\pdfoutput=1
\RequirePackage{ifpdf}
\ifpdf 
\documentclass[pdftex]{sigma}
\else
\documentclass{sigma}
\fi

\numberwithin{equation}{section}
\newtheorem{Theorem}{Theorem}[section]
\newtheorem*{Theorem*}{Theorem}

\newtheorem{Proposition}[Theorem]{Proposition}

\theoremstyle{definition}
\newtheorem{Definition}[Theorem]{Definition}

\newtheorem{Example}[Theorem]{Example}
\newtheorem{Remark}[Theorem]{Remark}

\usepackage{textcomp}
\usepackage{stmaryrd}

\newcommand{\changed}[1]{{#1}}
\renewcommand{\tilde}{\widetilde}

\renewcommand{\mod}{\,\rm mod \,}
\newcommand{\p}[1]{|#1|}
\newcommand{\gh}[1]{\mathrm{gh}(#1)}

\newcommand{\dx}{\mathrm{d}_X}

\renewcommand{\d}{\partial}
\renewcommand{\dh}{\mathrm{d_h}}

\newcommand{\cF}{\mathcal{F}}

\renewcommand{\geq}{\,{\geqslant}\,}
\renewcommand{\leq}{\,{\leqslant}\,}

\newcommand{\binner}[2]{%
 {\langle}\kern-4.15pt{\langle}#1{,}\,#2{\rangle}\kern-4.15pt{\rangle}}
\newcommand{\commut}[2]{[#1{,}\,#2]}

\newcommand{\half}{\mathchoice{%
 \ffrac{1}{2}}{\frac{1}{2}}{\frac{1}{2}}{\frac{1}{2}}}

\newcommand{\ffrac}[2]{\raisebox{.5pt}%
 {\footnotesize$\displaystyle\frac{#1}{#2}$}\kern1pt}

\newcommand{\dl}[1]{\mathchoice{\ffrac{\d}{\d #1}}{\frac{\d}{\d #1}}{\ffrac{\d}{\d #1}}{\ffrac{\d}{\d #1}}}

\newcommand{\Liealg}{\mathfrak}
\newcommand{\algg}{\Liealg{g}}

\newcommand{\algA}{\mathcal{A}}

\newcommand{\cC}{\mathcal{C}}

\newcommand{\fR}{\mathbb{R}}

\newcommand{\fZ}{\mathbb{Z}}

 \def\cD{\mathcal{D}}
 \def\cE{\mathcal{E}}

\def\cK{\mathcal{K}}

 \def\cN{\mathcal{N}}

\newcommand{\total}{\textbf{tot\,}}
\newcommand{\lb}{\llbracket}
\newcommand{\rb}{\rrbracket}

\newcommand{\vprol}{\tilde}

\begin{document}

\allowdisplaybreaks

\newcommand{\arXivNumber}{2408.08287}

\renewcommand{\PaperNumber}{096}

\FirstPageHeading

\ShortArticleName{Weak Gauge PDEs}

\ArticleName{Weak Gauge PDEs}

\Author{Maxim GRIGORIEV~$^{\rm a}$ and Dmitry RUDINSKY~$^{\rm bc}$}

\AuthorNameForHeading{M.~Grigoriev and D.~Rudinsky}

\Address{$^{\rm a)}$~Service de Physique de l'Univers, Champs et Gravitation, \\
\hphantom{$^{\rm a)}$}~Universit\'e de Mons, 20 place du Parc, 7000 Mons, Belgium}
\EmailD{\mail{grigoriev.max@gmail.com}}

\Address{$^{\rm b)}$~Institute for Theoretical and Mathematical Physics, \\
\hphantom{$^{\rm b)}$}~Lomonosov Moscow State University, 119991 Moscow, Russia}
 \EmailD{\mail{dmitry.rudinsky@yandex.ru}}
 \Address{$^{\rm c)}$~Lebedev Physical Institute, 53 Leninsky Ave., 119991 Moscow, Russia}

\ArticleDates{Received March 17, 2025, in final form November 03, 2025; Published online November 13, 2025}

\Abstract{Gauge PDEs generalise the AKSZ construction when dealing with generic local gauge theories. Despite being very flexible and invariant, these geometrical objects are usually infinite-dimensional and are difficult to define explicitly, just like standard infinitely-prolonged PDEs. We propose a notion of a weak gauge PDE in which the nilpotency of the BRST differential is relaxed in a controllable way. In this approach a nontopological local gauge theory can be described in terms of a finite-dimensional geometrical object. Moreover, among the equivalent weak gauge PDEs describing a given system, a minimal one can usually be found and is unique in a certain sense. In the case of a Lagrangian system, the respective weak gauge PDE naturally arises from its weak presymplectic formulation. We prove that any weak gauge PDE determines the standard jet-bundle Batalin--Vilkovisky formulation of the underlying gauge theory, giving an unambiguous physical interpretation of these objects. The formalism is illustrated by a few examples, including the non-Lagrangian self-dual Yang--Mills theory and a finite jet-bundle. We also discuss possible applications of the approach to the characterisation of those infinite-dimensional gauge PDEs that correspond to local theories.}


\Classification{35B06; 58A50; 37K06; 81T70; 70S15}

\section{Introduction}

Batalin--Vilkovisky (BV) formalism~\cite{Batalin:1981jr,Batalin:1983wj} gives an elegant way to uplift gauge systems to supergeometry. More precisely, a Lagrangian gauge system is embedded into a $\fZ$-graded symplectic supermanifold equipped with a compatible homological vector field. The non-Lagrangian version of this lift is obtained by forgetting the symplectic structure, see~\cite{Barnich:2004cr,Kazinski:2005eb,Sharapov:2016sgx}. If one is interested in local gauge theories and insists on maintaining manifest locality, the way out is the jet-bundle extension of the BV formulation~\cite{Barnich:1995ap, Barnich:1995db}, see also~\cite{Barnich:2000zw,Henneaux:1990rx,Piguet:1995er}, which gives a powerful framework to analyse anomalies, consistent interactions, and renormalization in terms of local BRST cohomology.

Despite being a powerful framework, the jet-bundle BV approach is not sufficiently flexible and hides the geometry of the underlying gauge fields and their symmetries. For instance, it is not very well suited to study the relation between gauge fields in the bulk and their values at~submanifolds/boundaries. A more invariant version of the manifestly local BV formulation is the so-called AKSZ construction~\cite{Alexandrov:1995kv} (see also~\cite{Barnich:2009jy,Batalin:2001fc,Bonavolonta:2013mza,Bonechi:2009kx,Cattaneo:1999fm,Cattaneo:2001ys,Grigoriev:1999qz,Ikeda:2012pv,Roytenberg:2002nu}) that was initially developed in the context of topological theories. In particular, AKSZ sigma-models behave well when restricted to submanifolds/boundaries~\cite{Barnich:2003wj,Barnich:2006hbb,Cattaneo:2012qu,Cattaneo:2015vsa, Grigoriev:1999qz}. The generalisation of the non-Lagrangian version of AKSZ sigma-models to generic local gauge theories is also known under the name of gauge PDEs~\cite{Grigoriev:2019ojp}, see also~\cite{Barnich:2010sw, Barnich:2004cr}, where these objects were initially introduced in a less general and less geometrical framework. Gauge PDEs can also be understood as a BV-BRST extension of the unfolded approach~\cite{Vasiliev:1988xc,Vasiliev:2005zu} of higher spin theory.
Gauge PDEs (a gPDE, for short) are supergeometrical objects that encode the BV description of local non-Lagrangian gauge theories. Particular cases of gPDEs include usual PDEs, jet-bundle BV formulations, AKSZ sigma-models, and free differential algebras. Just like AKSZ sigma-models, gPDEs behave well when restricted to submanifolds, which makes them useful~\cite{Bekaert:2013zya,Bekaert:2012vt,Bekaert:2017bpy,Chekmenev:2015kzf,Grigoriev:2023kkk} in the study of gauge theories on manifolds with boundaries. However, gPDEs are infinite-dimensional unless the underlying system is a PDE of finite type or a
topological theory. Moreover, just like usual PDEs, gauge PDEs are often defined only implicitly as infinite prolongations of subbundles of jet-bundles. Of course, one can always define a local gauge theory in terms of a finite jet-bundle, but such a description is far from being elegant and is very ambiguous. The ambiguity arises, in particular, from the fact that the same gauge theory can be formulated using different field content. For example, the first-order and second-order formulations of a given physical system employ distinct sets of fields.

An interesting alternative, advocated in this work, is to consider a finite-dimensional analog of a gPDE in which the BRST differential $Q$ is nilpotent only modulo certain integrable $Q$-invariant distribution $\cK$. In particular, this structure naturally arises from the weak presymplectic formulation~\cite{Dneprov:2024cvt, Grigoriev:2022zlq}, see also~\cite{Alkalaev:2013hta,Dneprov:2022jyn,Grigoriev:2016wmk,Grigoriev:2020xec,Sharapov:2021drr}, of Lagrangian systems. More specifically, the distribution $\cK$ is a suitable version of the kernel distribution of the presymplectic structure. Another perspective on weak gPDEs is that they arise as certain truncations of genuine gPDEs. In particular, starting from a minimal gPDE describing a given gauge system, one can consider its maximal consistent truncation as a sort of minimal weak gPDE that describes the system, see~\cite{Dneprov:2024cvt} for a more detailed discussion of such minimal models in the case of Lagrangian systems.\looseness=1

The main statement proved in this work is that any weak gPDE naturally determines a~jet-bundle BV formulation defined on a certain quotient of the super-jet bundle of the initial weak gPDE. This can be considered as a far-reaching generalisation of the AKSZ construction. Indeed, the non-Lagrangian version~\cite{Barnich:2006hbb,Grigoriev:2006tt}, see also~\cite{Barnich:2009jy, Kotov:2007nr}, of the AKSZ construction determines a~local BV system in terms of the trivial bundle over $T[1](\text{spacetime})$ and the
fiber being the $\fZ$-graded $Q$-manifold. Now, the analogous geometrical input, in which the bundle is allowed to be nontrivial and $Q^2$ belongs to the distribution, still defines a local BV system that is not necessarily topological or diffeomorphism-invariant.

The paper is organised as follows: in Section~\ref{sec:prelim}, we recall the basic notions of $Q$-manifolds, non-Lagrangian BV formalism on jet-bundles and gauge PDEs. In the main Section~\ref{sec:wgpde}, we define weak gPDEs and prove the main Theorem~\ref{prop:main}. In addition, in this Section we introduce auxiliary notions needed to work with weak gPDEs. These include vertical jets and weak local BV systems.
In the final Section~\ref{sec:examples}, we collected a few explicit examples. The first one, the scalar field, illustrates the construction of this work in the case of a system without gauge freedom. The second is the minimal weak gPDE formulation of the self-dual Yang--Mills gauge theory which is a genuine non-Lagrangian gauge system. \changed{We also use this example to demonstrate that the weak gPDE formulation naturally leads to the AKSZ-like formulation of the type proposed in~\cite{costello2011renormalization} by K.~Costello, see also~\cite{Bonechi:2022aji,Bonechi:2009kx,Grigoriev:2025vsl}, and based on the non-freely generated differential graded commutative algebra which replaces the space-time exterior algebra.} As a third example we demonstrate that a finite-order jet bundle naturally gives rise to a weak gPDE whose inequivalent solutions are one-to-one with the sections of the initial bundle. In the concluding section, we discuss possible applications of the construction.\looseness=1

\section{Preliminaries}
\label{sec:prelim}
\subsection[Q-manifolds and gauge systems]{$\boldsymbol{Q}$-manifolds and gauge systems}

Throughout the paper, we utilize the language of $Q$-manifolds and related supergeometrical objects. Below we recall some basic definitions and examples.
More details and original references can be found, e.g., in \cite{Kotov:2007nr,Mehta:2007rgt,Schwarz:1992nx,Vaintrob:1997,Voronov:2009nr}.
\begin{Definition}
 A $Q$-manifold is a pair $(M,Q)$, where $M$ is a $\mathbb{Z}$-graded supermanifold and $Q$ is an odd vector field of degree $1$, such that $[Q,Q]=0$. Such vector field is called homological.
\end{Definition}
The $\fZ$-degree is called ``ghost degree'' and is denoted by $\gh{\cdot}$ and the Grassmann parity by~$\p{\cdot}$. To simplify the exposition, we assume that the two degrees are compatible, i.e., ${\p{\cdot}=\gh{\cdot} \mod 2}$.
From the field theory point of view, this means that there are no physical fermions. To incorporate fermionic degrees of freedom and fermionic gauge symmetries, one should relax the requirement that the ghost degree and the Grassmann degree are compatible. This is a~standard and straightforward generalization. We~refer to~\cite{Grigoriev:2025vsl} for a treatment of supersymmetric systems in the closely related framework.

The very basic and useful example of $Q$-manifold is as follows.

\begin{Example}
 Given a (graded) manifold $M$, let $T[1]M\rightarrow M$ be its shifted tangent bundle. $T[1]M$ is a $Q$-manifold, with the $Q$-structure given by the de Rham differential \smash{$\mathrm{d_M}=\theta^a\frac{\partial}{\partial x^a}$}, where $x^a$, $a=1,\dots, \dim(M)$
 are base coordinates and $\theta^a$ are fiber ones of degree $1$.
\end{Example}

Morphisms of $Q$-manifolds are maps between the underlying graded supermanifolds that preserve the $Q$-structure.

\begin{Definition}
 Let $(M_1,Q_1)$ and $(M_2,Q_2)$ be two $Q$-manifolds. A degree preserving map $\phi\colon M_1\rightarrow M_2$ is called a $Q$-morphism (or a $Q$-map) if it satisfies the following condition:
 \begin{equation*}
 Q_1\circ\phi^*=\phi^*\circ Q_2,
 \end{equation*}
 where $\phi^*$ is a pullback of $\phi$.
\end{Definition}

$Q$-manifolds provide a useful geometrical description of gauge systems at the level of equations of motion. Let us recall how the data of a gauge system in $0$ dimensions is encoded in a given $Q$-manifold.

Let $(M,Q)$ be a $Q$-manifold that encodes a gauge system and $(pt,0)$ denotes a trivial $Q$ manifold, where $pt$ is just a point. Solutions of the underlying gauge system are identified with $Q$-morphisms $\sigma\colon pt\rightarrow M$, i.e., maps satisfying $\sigma^*\circ Q=0$. In other words, solutions are points of the zero locus of $M$. The infinitesimal gauge transformation of a given map $\sigma$ is defined as
\begin{equation*}
 \delta\sigma^*=\sigma^*\circ[Q,Y],
\end{equation*}
where $Y$ is a vector field on $M$ of degree $-1$, referred to as the gauge parameter. It is easy to check that the above gauge transformation transforms solutions to solutions. \changed{Gauge for gauge symmetries, which are present if the gauge transformations are redundant, can be defined analogously.}\footnote{\changed{By definition, redundant gauge transformations of gauge parameters that do not affect the transformation of fields, see, e.g.,~\cite{Gomis:1995he, HT-book} for further details.}} For instance, $\delta Y=\bigl[Q,Y^{(1)}\bigr]$, where ``gauge for gauge parameter'' $Y^{(1)}$ is a vector field of degree $-2$.

\subsection{Jet-bundle BV description}

A systematic framework to describe local gauge theories maintaining manifest locality is provided by the jet-bundle BV approach. The detailed exposition of the approach in the case of Lagrangian theories is available, e.g., in~\cite{Barnich:2000zw}. Now we are concerned with gauge theories at the level of equations of motion. A suitable extension of the jet-bundle BV to this case is achieved by forgetting the BV symplectic structure. The following definition is a more geometrical version of the one from~\cite{Barnich:2004cr} (see also~\cite{Barnich:2006hbb,Barnich:2010sw,Grigoriev:2022zlq,Kazinski:2005eb}).

\begin{Definition}
\quad
\begin{itemize}\itemsep=0pt
\item
 A local non-Lagrangian BV system is a graded fiber bundle $\mathcal{E}\rightarrow X$ together with a~degree $1$ evolutionary vector field $s$ on an infinite jet-bundle $J^{\infty}\mathcal{E}$, which is required to be homological, i.e., \smash{$s^2=\half\commut{s}{s}=0$}.

 \item Section $\sigma\colon X\rightarrow \mathcal{E}$ is called a solution if its jet prolongation $\bar{\sigma}\colon X\rightarrow J^{\infty}\mathcal{E}$ satisfies
 \begin{equation}\label{bveom}
 \bar{\sigma}^*\circ s=0.
 \end{equation}

\item Gauge parameters of level $l=0,1,2,\dots$ are vertical vector fields on $\mathcal{E}$ of ghost degree~${-1-l}$. Gauge parameters of level $0$ are referred to as gauge parameters.

\item Infinitesimal gauge transformations are defined by
 \begin{equation*}
 \delta\bar{\sigma}^*=\bar{\sigma}^*\circ\bigl[s, \bar Y\bigr],
 \end{equation*}
 where $\bar{Y}$ is a jet-prolongation of the gauge parameter vector field $Y$. Level $l$, $l>0$ gauge transformation is given by
 \begin{equation*}
 \delta Y^{(l)}=\bigl[s,Y^{l+1}\bigr].
 \end{equation*}
\end{itemize}
\end{Definition}
Here and in what follows, $J^{\infty}\cE$ denotes a superjet-bundle (jets of supersections in contrast to jets of sections, see, e.g.,~\cite{bonavolonta2013}). In particular, $J^{\infty}\cE$ is necessarily a graded manifold if the grading of $\cE$ is nontrivial. Equation \eqref{bveom} says that section $\sigma$ is a solution if its prolongation $\bar\sigma$ belongs to the zero-locus of $s$. Note that $\bar{\sigma}$ is a map of graded manifolds and hence for an arbitrary function $f\in C^{\infty}(J^{\infty}\mathcal{E})$ of nonvanishing degree, one has $\bar{\sigma}^*(f)=0$, because on $X$ there are no functions of non-vanishing degrees .

Because in this work we are only concerned with gauge systems at the level of equations of motion, we systematically omit the term ``non-Lagrangian'' in the names of objects. For instance, ``local non-Lagrangian BV systems'' defined above are refereed to in what follows as ``local BV systems'', unless otherwise specified.

\subsection{Gauge PDEs}
Although local BV systems provide a powerful framework to study local gauge theories, it is not sufficiently flexible and geometrical. A more flexible framework is provided by so-called gauge PDEs.

\begin{Definition}\quad
\begin{itemize}\itemsep=0pt
 \item A $\mathbb{Z}$-graded fiber bundle $\pi\colon E \rightarrow T[1]X$ equipped with a homological vector field $Q$: $\gh{Q} =1$, $[Q,Q]=0$ such that $Q\circ\pi^* = \pi^*\circ \dx$ is called a gauge PDE (gPDE) and is denoted $(E,Q,T[1]X)$, where $\dx$ is the de Rham differential understood as a vector field on $T[1]X$.
 \item Sections of $E$ are interpreted as field configurations. Section $\sigma\colon T[1]X\rightarrow E$ is called a~solution if\footnote{The equation \eqref{geom} represents the condition that the zeros of a vector field $\mathcal{R}_\sigma:=\dx\circ \sigma^* - \sigma^* \circ Q$ along the map $\sigma$ are precisely those $\sigma$ for which $\dx$ and $Q$ are $\sigma$-related.}
 \begin{equation}\label{geom}
 \dx\circ \sigma^*=\sigma^*\circ Q,
 \end{equation}
 where $\sigma^*$ is a pullback of $\sigma$.
 \item Infinitesimal gauge transformations of a given section $\sigma$ are defined by
 \begin{equation}\label{gt}
 \delta\sigma^*:=\sigma^*\circ [Q,Y],
 \end{equation}
 where $Y$ is a vertical vector field of degree $-1$ and it is understood as a gauge parameter. In a similar way, one defines gauge for gauge symmetries.
\end{itemize}
\end{Definition}
In addition, one often requires $(E,Q,T[1]X)$ to be equivalent to a local BV system in order to exclude nonlocal theories. Another natural requirement is the equivalence to a nonegatively graded gPDE. See~\cite{Grigoriev:2019ojp} for the precise notion of equivalence of gPDEs and further details.

It is easy to check that~\eqref{gt} preserves the equation \eqref{geom}. In a similar way, thanks to~\eqref{gt}
the gauge for gauge transformation $\delta Y=\bigl[Q,Y^{(1)}\bigr]$ of gauge parameter $Y$ does not affect the gauge transformation it determines. It should be noted that the gauge transformation of \eqref{geom} can also be defined as follows:
\begin{equation*}
 \delta\sigma^*:=\dx\circ\xi^*_{\sigma}+\xi^*_{\sigma}\circ Q,
\end{equation*}
where $\xi^*_{\sigma}\colon C^{\infty}(E)\rightarrow C^{\infty}(T[1]X)$ is a degree $-1$ vector field along $\sigma$, i.e., $\xi^*_{\sigma}$ satisfies to the graded version of \eqref{vam},
\begin{equation*}
 \xi^*_{\sigma}(fg)=\xi^*_{\sigma}(f)\sigma^*(g)+(-1)^{|f|}\sigma^*(f)\xi^*_{\sigma}(g),
\end{equation*}
where $f,g\in C^{\infty}(E)$. Furthermore, $\xi^*_{\sigma}$ satisfies the condition $ \xi^*_{\sigma}(\pi^*\alpha)=0$ for all $\alpha\in C^{\infty}(T[1]X) $, which excludes gauge transformations involving reparametrisations of $X$. If one takes $\xi^*_{\sigma}=\sigma^*\circ Y$, then one indeed recovers \eqref{gt} if $\sigma$ is a solution.

It is easy to see that a local BV system gives rise to a gauge PDE of special form. Namely, pulling back $J^\infty\cE$ from $X$ to $T[1]X$ one can identify functions on the total space $E$ with horizontal forms on $J^\infty\cE$ so that the horizontal differential $\dh$ on $J^\infty\cE$ can be seen as a homological vector field. In this way, we arrive at $Q$-bundle $E$ over $T[1]X$, whose $Q$-structure is given by~${Q=\dh+s}$. It can be proved~\cite{Barnich:2010sw} (see also~\cite{Barnich:2004cr,Grigoriev:2019ojp}) that the resulting gauge PDE is equivalent to the starting point local BV system.

\section{Weak gauge PDEs}\label{sec:wgpde}

Although gPDEs seem to be right geometrical objects underlying local gauge theories they are often not very practical because $E$ is finite-dimensional only in special cases such as mechanical systems, topological theories or PDEs of finite type. The situation here is analogous to the geometrical theory of PDEs~\cite{Krasil?shchik-Lychagin-Vinogradov,Krasil'shchik:2010ij, Vinogradov1981}, where the invariant geometrical object underlying a~given PDE (which can be seen as a special case of gPDE) is usually infinite-dimensional and in the case of nonlinear PDEs can be often defined only implicitly. As the example relevant in the gPDE context, let us mention that the explicit form of minimal gPDE formulation (such formulations are also known as unfolded) of Yang--Mills theory in four dimensions has appeared only recently~\cite{Misuna:2024dlx} and employs dimension-specific techniques.\footnote{Nevertheless, let us mention that, so far, the covariant formulation of chiral higher spin gravity is only known in the unfolded/gPDE terms~\cite{Sharapov:2022faa, Sharapov:2022wpz}. The same applies to AdS higher spin gravity~\cite{Vasiliev:1999ba, Vasiliev:2003ev} but this does not seem to determine a sensible perturbatively-local system.} It goes without saying, that such a formulation is by far more involved than the usual BV formulation of Yang--Mills. Of course, one can always reconsider the conventional jet-bundle BV formulation of a given system as a gPDE and hence achieve the explicit gPDE description but such a reformulation is quite ambiguous and is still infinite-dimensional.

In this section, we introduce a concept of weak gPDEs. These objects can be thought of as~specific finite-dimensional truncations of gPDEs that still contain all the information about the underlying gauge theory. In particular, the local BV system describing the underlying gauge theory can be constructed from a weak gPDE in a systematic way.

\subsection[Weak Q-manifolds]{Weak $\boldsymbol{Q}$-manifolds}

To illustrate the main idea of the approach, it is instructive to start with the 
case of gauge theory in 0 dimensions. In this case, a gPDE is just a $Q$-manifold.
The generalization amounts to relaxing the nilpotency condition $[Q,Q]=2Q^2=0$ in a controllable way. Namely, we let $Q^2$ to belong to an involutive distribution which is an additional geometrical structure present on the underlying manifold. As we are going to see, $Q$ is nilpotent on the subalgebra of functions annihilated by the distribution and hence under the standard technical assumptions makes the respective quotient into a $Q$-manifold. This gives a way to encode $Q$-manifolds in terms of more flexible structures.

The above idea is formalized as follows.

\begin{Definition}
 A weak $Q$-manifold is a $\mathbb{Z}$-graded manifold $M$ equipped with a degree-$1$ vector field $Q$ and a distribution $\mathcal{K}$ satisfying
\begin{equation}\label{cond1}
 [Q,Q]=2Q^2\in \mathcal{K},\qquad L_Q\mathcal{K}\subseteq \mathcal{K},\qquad [\mathcal{K},\mathcal{K}]\subseteq \mathcal{K}.
\end{equation}
\end{Definition}
Note that the distribution involved in the above definition is generally not required to be a~subbundle of $TM$. For the moment, we only require it to be a finitely-generated submodule of the $\cC^\infty(M)$-module $\mathfrak{X}(M)$ of vector fields on $M$. It is clear that $Q$ is defined modulo degree~$1$ vector field from~$\cK$ and hence can be naturally considered as an equivalence class modulo $\cK$.
\begin{Remark}
 More generally, instead of globally defined vector field $Q$ one can give representatives $Q_\alpha$ satisfying \eqref{cond1} in each coordinate patch and such that in the overlaps of patches $Q_\alpha-Q_\beta \in \cK$. In this case, the equivalence class of $\cK$ is still globally well-defined.
\end{Remark}

The following proposition illustrates in which sense weak $Q$-manifolds encode
genuine gauge systems.

\begin{Proposition}\label{prop1}
 Let $(M,Q,\mathcal{K})$ be a weak $Q$-manifold and let $\mathcal{A}\subset C^{\infty}(M)$ be a subalgebra of functions annihilated by $\mathcal{K}$. Then $Q$ preserves $\mathcal{A}$ and $Q^2(f)=0$ for any $f\in\mathcal{A}$.
\end{Proposition}
In particular, it follows that if the distribution $\mathcal{K}$ is regular, then, at least locally, the quotient of $M$ by $\mathcal{K}$ is a $Q$-manifold.

\begin{proof}
For any $f\in\mathcal{A}$ and $K\in\mathcal{K}$ one has $KQ(f)=\pm QK(f)=0$, where $\pm$ depends on the degree of $K$ and we made use of $\commut{Q}{K}\in \cK$. It follows $Q$ preserves $\mathcal{A}$ and hence $Q^2(f)=0$ thanks to $[Q,Q]\in \mathcal{K}$.
\end{proof}

A class of examples of weak $Q$-manifolds is provided by weak presymplectic $Q$-manifolds~{\cite{Dneprov:2024cvt, Grigoriev:2022zlq}} which are known to encode Lagrangian gauge systems. More precisely, we have the following.

\begin{Example}
 Let $M$ be equipped with a vector field $Q$, $\gh{Q}=1$ and a closed 2-form $\omega$, $\gh{\omega}=k$ satisfying $L_Q\omega=0$, $i_Qi_Q\omega=0$. It is not hard to show that conditions \eqref{cond1} are satisfied for $\cK$ being the kernel distribution of $\omega$ so that $(M,Q,\cK)$ is a weak-$Q$ manifold.
\end{Example}
Somewhat similar idea underlies the following construction, where the distribution is defined as the kernel of a~$(1,1)$ tensor field $P$.
\begin{Example} \label{ex:projector}
 Let $M$ be equipped with a vector field $Q$, $\gh{Q}=1$. Let us endow $M$ with a~$(1,1)$ tensor field $P$ of degree $\gh{P}=k$, which satisfies
 \begin{equation*}
 P^2=P, \qquad L_QP=0,\qquad L_Q(i_QP)=0, \qquad \cN_P=0,
 \end{equation*}
 where $ \cN_P$ is the Nijenhuis tensor of $P$. Let $\mathcal{K}$ be a kernel distribution of $P$. It is a well-known fact (see, e.g., \cite{Kobayashi:1962a}) that $\cN_P=0$ and $P^2=P$ imply that $\mathcal{K}$ is involutive. Moreover, $L_QP=0$ implies that $\cK$ is preserved by $Q$. Finally, $L_Q(i_QP)=0$ implies that $Q^2$ lies in $\mathcal{K}$ so that we are dealing with a weak $Q$-manifold.
\end{Example}

\begin{Remark}
 It is useful to briefly discuss $L_{\infty}$-structures associated to formal weak $Q$-manifolds. Recall that formal pointed $Q$-manifolds are $1:1$ with $L_\infty$-algebras \changed{see, e.g.,~\cite{Alexandrov:1995kv,Kontsevich:1997vb}}. More precisely, $Q$ determines multilinear operation on the underlying vector space $L$ while the nilpotency condition $Q^2=0$ is equivalent to higher Jacobi identities satisfied by them. Let us consider a~weak $Q$-manifold $(M,Q,\mathcal{K})$ which we assume formal. We also assume $\cK$ regular so that in a~suitable coordinate system it is constant and hence can be identified with the subspace $K\subset L$. It is not difficult to see that $Q^2\subset \cK$ implies that the higher Jacobi identities are satisfied modulo~${K\subset L}$. In particular, the quotient space $L/K$ is an $L_\infty$ algebra. These consideration can be extended to weak gPDEs following the lines of~\cite{Grigoriev:2023lcc}.
\end{Remark}

\begin{Remark}
An alternative interpretation of weak $Q$-manifolds can be given in terms somewhat generalised Lie algebroids. More precisely, suppose that our weak $Q$-manifold has coordinates of degree $0$ and $1$ only. If the distribution $\cK$ is trivial, then according to~\cite{Vaintrob:1997} we are dealing with a Lie algebroid. More precisely, the manifold itself can be identified with $E[1]X$ where~${E \to X}$ is the vector bundle underlying our graded manifold. In so doing, all the structures of the Lie algebroid are encoded in $Q$. If $\cK$ is nontrivial, let us for simplicity assume that it is generated by the vector fields of degree $0$. In this case, $\cK$ is clearly tangent to $X \subset E[1]X$ (embedded as a zero section of $E$) and defines an involutive distribution $\cK_0$ therein.
By inspecting the condition $Q^2\in \cK$,
one finds that the axioms of the underlying Lie algebroid are satisfied modulo the distribution $\cK_0$. For instance, choosing local coordinates $x^i$ and local trivialisation~$e_\alpha$
one finds that the condition that the anchor is a homomorphism is fulfilled modulo~$\cK_0$. If~$\cK$ is regular, then the (possibly locally defined) quotient of $E[1]X$ by $\cK$ is already a~standard Lie algebroid. Analogous considerations can be applied to more general nonnegatively graded $Q$-manifolds and their associated Lie $n$-algebroids, see, e.g.,~\cite{Bonavolonta:2012fh,Severa:2001tze} for more details on Lie $n$-algebroids.
\end{Remark}

\subsection{Vector fields along sections (maps)}

In what follows, we need a notion of vector fields along maps. Here we recall the notion in the form suitable for supermanifolds.
\begin{Definition}
 Let $\phi\colon M_1\rightarrow M_2$ be a smooth map. A vector field $V_{\phi}$ along $\phi$ is a derivation of functions on $M_2$ with values in functions on $M_1$
 \begin{equation*}
 V_{\phi}\colon\ C^{\infty}(M_2)\rightarrow C^{\infty}(M_1),
 \end{equation*}
 for which the following Leibniz-type property holds:
 \begin{equation}\label{vam}
 V_{\phi}(fg)= V_{\phi}(f)\phi^*(g)+(-1)^{\p{V}\p{f}}\phi^*(f) V_{\phi}(g), \qquad \forall f,g\in \cC^{\infty}(M_2).
 \end{equation}
\end{Definition}
Alternatively, a vector field along $\phi\colon M_1\rightarrow M_2$ is a section of a pullback bundle $\phi^*(TM_2)\rightarrow M_1$, see, e.g.,~\cite{Lee2013}.

It is clear that vector field along $\phi\colon M_1\rightarrow M_2$ form a module over $\cC^{\infty}(M_1)$. One can also define a~$\cC^{\infty}(M_2)$-module structure by taking $(fV)(g)=\phi^*(f)V(g)$ for all $f\in \cC^{\infty}(M_2)$. Any vector field $V$ on~$M_2$ gives rise to a vector field $\phi^*\circ V$ along $\phi$.

One can also speak of distributions over $\phi$. These are submodules of all vector fields along~$\phi$ seen as a~module over $\cC^\infty(M_1)$. In particular, any distribution $\cK$ on $M_2$ gives rise to a~distribution~${\phi^*\circ \cK}$ along $\phi$, which is generated by vector field along $\phi$ of the form $\phi^*\circ K$, $K\in \cK$. Note that if $V$ and $V^\prime$ coincide at $\phi(M_1)$, then $\phi^*\circ V=\phi^*\circ V^\prime$.

Let us prove the following useful proposition.

\begin{Proposition}\label{prop2}
 Let $\phi\colon M_1\rightarrow M_2$ be an embedding
 and $M_2$ be equipped with a distribution $\mathcal{K}$. For a vector field $V$ along $\phi$, the following conditions are equivalent:
 \begin{enumerate}\itemsep=0pt
 \item[$(i)$] $V$ belongs to $\phi^*\circ \cK$.
 \item[$(ii)$] $Vf=0$ for any $($possibly locally defined$)$ function annihilated by $\cK$.
 \end{enumerate}
\end{Proposition}
\begin{proof}
The statement is local so we restrict ourselves to a contractible domain. First of all note that under the conditions, any $V$ along $\phi$ can be represented as $\phi^*\circ V^\prime$ for some vector field~$V^\prime$ on~$M_2$. Then~(ii) implies $0=Vf=(\phi^*\circ V^\prime) f=\phi^*(V^\prime f)$ so that $(V^\prime f)$ vanishes on $\phi(M_1)$ and hence $V^\prime|_p\in \cK|_p$ for all $p \in \phi(M_1)$. It follows $V$ belongs to $\phi^*\circ \cK$.
Other way around, (i) means that $V=v^\alpha (\phi^*\circ K_\alpha)$ for some $v^\alpha \in\cC^{\infty}(M_1)$ and $K_\alpha \in \cK$, giving (ii).
\end{proof}

\subsection{Vertical jets} \label{sec:vert-jets}
\changed{Let us briefly recall the construction of jet-bundles, for further details see, e.g.,~\cite{Andersonbook, Vinogradov1981}. Given a fibre bundle $\mathcal{E}\rightarrow X$ a fibre of the order-$k$ jet bundle $J^k\mathcal{E} \rightarrow X$ at $x \in X$ is given by the space of equivalence classes of sections (possibly locally-defined) with respect to the following equivalence relation: sections $\phi$, $\psi$ are equivalent if all their partial derivatives to order $k$ coincide at $x$. There is a natural projection $J^{k+1}\cE \to J^{k}\cE$ and $J^0\cE=\cE$.} \changed{The infinite jet bundle~$J^{\infty}\mathcal{E}$ is then defined as the inverse (also known as projective) limit of the chain of projections
\[
\dots \rightarrow J^k\mathcal{E} \rightarrow J^{k-1}\mathcal{E} \rightarrow \dots \rightarrow \mathcal{E}.
\]}

Another ingredient which we need in what follows is the vertical jet-bundle. The setup is as follows: let $\pi\colon E\rightarrow B$ be a fibre bundle whose base $B$ is itself a bundle $p\colon B\rightarrow X$ over $X$. In analogy to the standard definition of $k$-jets one can now introduce the following equivalence relation: two (locally defined) sections $B\to E$ are equivalent at $b\in B$ if all their derivatives of order $\leq k$ along vertical directions in $B$ coincide at $b$. It is easy to check that the equivalence does not depend on the choice of coordinates and defines a fibre of the vertical jet-bundle \changed{over a point $b\in B$, which we denote as $J_V^kE$}. The construction can be generalised to the case where the base $B$ is equipped with a foliation which is not necessarily a fibration.

Although the above construction is very intuitive, it is difficult to literally apply it to the case of supermanifolds. In the case, where $X$ is a real manifold, one can easily define vertical jets without the explicit reference to the equivalence classes of sections. In what follows, we assume that $X$ is a real manifold while $E$ and $B$ are generally graded manifolds. Unless otherwise specified, by jets we always mean jets of super-sections (not to be confused with jets of sections).

For each point $x\in X$, let $B_x$ denote a fiber of $B$ over $x$. Then we have a natural fiber bundle $\pi_x\colon E_x\rightarrow B_x$, where $E_x\equiv E|_{B_x}$. Given the fiber bundle $E_x \rightarrow B_x$, its associated infinite jet-bundle $J^{\infty}E_x$ is constructed in a standard way. Finally, taking a disjoint union of $J^{\infty}E_x$ over~$X$ gives a well-defined bundle $J^{\infty}_VE\rightarrow B$ over $B$. Moreover, $J^{\infty}_VE$ is a well-defined bundle over~$E$ and we denote it by $\pi^{\infty}_V\colon J^{\infty}_VE\rightarrow E$.
Furthermore, any section $\sigma\colon B\rightarrow E$ has a canonical vertical prolongation to a section of $J^{\infty}_VE\rightarrow B$ which can be defined through the standard prolongations of $\sigma$ restricted to $B_x$.

In a similar way, one generalizes to vertical jets the concept of total vector fields. Namely, if~$V$ is a vertical vector field on $B$, its associated total vector field $\total{V}$ is the vector field on $J^{\infty}_VE$ that projects to $V$ and for any section $\sigma$ and a local function $f$ satisfies
\begin{equation*}
 \vprol{\sigma}^*(\total{V}f)=V (\vprol{\sigma}^* (f)),
\end{equation*}
where $\vprol{\sigma}$ is a vertical prolongation of $\sigma$ and $\vprol{\sigma}^*$ its associated pullback map. In particular, if~$x^\mu$,~$v^a$ are adapted local coordinates on $B$ such that $x^\mu$ are pullbacks of local coordinates on $X$, one~can define vertical total derivatives \smash{$D^{(v)}_a$} as follows: \smash{$D^{(v)}_a=\total{\dl{v^a}}$}.

The analog of the usual prolongation of vertical vector fields can be defined as follows.

\begin{Definition}
 Let $V$ be a vertical vector field on $E$. Its vertical prolongation to $J^{\infty}_VE$ is the vector field $\vprol{V}$ on $J^{\infty}_VE$ such that
 \begin{enumerate}\itemsep=0pt
 \item[(1)] $(\pi^{\infty}_E)_*\bigl(\vprol{V}\bigr)=V$,
 \item[(2)] $\bigl[D^{(v)}_a,\vprol{V}\bigr]=0$,
 \end{enumerate}
 where $(\pi^{\infty}_V)_*\colon TJ^{\infty}_VE\rightarrow TE$ denotes the differential of the projection $\pi^{\infty}_E$.
\end{Definition}

In what follows, we are mainly concerned with the fibre bundles over $T[1]X$, with $X$ being a real manifold, so that the vertical jet-bundle $J_V^\infty E$ is finite dimensional provided $E$ is. It is convenient to introduce a new bundle $\bar E \to X$ to be the restriction of $J^\infty_V E$ to the zero section of $T[1]X\to X$. In this setup, it is easy to see that the fiber of $\bar E$ at $x\in X$ is the supermanifold of supermaps from $T_x[1]X$ to the fiber of $E$.
\begin{Remark}
The jet-bundle $J^\infty E$ restricted to the zero section of $T[1]X\to X$ coincides with~${J^\infty \bar E}$.
\end{Remark}
Another useful property of vertical jets over $T[1]X$ is the following.

\begin{Remark}\label{rem:pullback}
$J_V^\infty E \to T[1]X$ ($J^\infty E\to T[1]X$) is a pullback of $\bar E \to X$ \big(resp.\ $J^\infty \bar E \to X$\big) by the canonical projection $T[1]X\to X$.
\end{Remark}
Later in this section we show that this is indeed true.

Finally, let us sketch the coordinate description of the vertical jets of $E\rightarrow T[1]X$. Let $\bigl(x^a,\theta^a,\psi^A\bigr)$ be local coordinates on $E\rightarrow T[1]X$ adapted to local trivialization. Vertical total derivatives
are denoted by \smash{$D^{(\theta)}_a=\total{\dl{\theta^a}}$}. A convenient coordinate system on $J^{\infty}_VE$ is given by \smash{$\bigl(x^a,\theta^a,\bar\psi^A,\bar\psi^A_{|a_1\dots}\bigr)$}, where \smash{$\bar\psi^A_{|a_1\dots a_k}$} is totally antisymmetric with respect to the lower indices,
\[
{\rm gh}\bigl(\bar\psi^A_{|a_1\dots a_k}\bigr)={\rm gh}\bigl(\psi^A\bigr)-k.
\]
 The additional conditions defining this coordinate system is that~\smash{$D^{(\theta)}_a=\dl{\theta^a}$} and the canonical projection $\pi_V^\infty:J^\infty_VE\to E$ reads as
\begin{equation}
\label{projection-vert-jets}
(\pi_V^\infty)^*\bigl(\psi^A\bigr)=\bar\psi^A+\theta^a \bar\psi^A_{|a}+\half \theta^a\theta^b\bar\psi^A_{|ab}+\cdots.
\end{equation}
Somewhat implicitly the construction of vertical jets in this case was in
\cite{Grigoriev:2022zlq, Grigoriev:2019ojp,Grigoriev:2020xec} (see also~\mbox{\cite{Barnich:2009jy,Barnich:2010sw,bonavolonta2013}}).

In the above coordinates it is easy to see that Remark~\ref{rem:pullback} is true. Let $x\in X$ and $({J^\infty_VE})_x$ denote a restriction of $J^\infty_VE$ to $T_x[1]X$. Consider the horizontal distribution generated by~\smash{$D^{(\theta)}_a$}.~It determines a fibration because it is integrable and purely odd. Moreover, it is transversal to the fibres of ${(J^\infty_V E)}_x$ and hence ${(J^\infty_V E)}_x$ can be identified as a product of the fiber and $T_x[1]X$. Because the fiber can be seen as a restriction of ${(J^\infty_V E)}_x$ to the zero point of $T_x[1]X$, it coincides with the fiber of $\bar E$ at $x$ and hence one concludes that ${(J^\infty_V E)}_x$
is a pullback of $\bar E_x$ by the projection $T_x[1]X \to 0$. Applying the same argument for all $x\in X$ shows that $J^\infty_V E$ can be identified with a pullback of $\bar E$ by $T[1]X\to X$.

Let us also give an example of what the vertical prolongation of a vector field looks like in coordinates.

\begin{Example}
 Let $E$ be a trivial bundle over $T[1]\fR^1$ with the fiber $\fR^1$. The standard adapted coordinate system is $x$, $\theta$, $\psi$. Consider a vector field \smash{$V=\theta \dl{\psi}$}. The associated coordinate system on~$J^\infty_V E$ is given by $x$, $\theta$, $\bar\psi_0$, $\bar\psi_1$ and the projection acts as $(\pi^\infty_V)^*(\psi)=\bar\psi_0+\theta \bar\psi_1 $. In this coordinate system, the prolongation \smash{$\vprol V$} of $V$ is given by \smash{$\vprol V=-\frac{\partial}{\partial \bar{\psi}_1}$}.
\end{Example}

\begin{Remark}
The important observation is that the vertical prolongation of non-regular vector field can be regular.
\end{Remark}
The above observation motivates the following definition.

\begin{Definition}
 Let $\mathcal{V}$ be a vertical distribution on $E$ and \smash{$\vprol{\mathcal{V}}$} be its vertical prolongation to~$J^{\infty}_VE$. $\mathcal{V}$ is called quasi-regular if \smash{$\vprol{\mathcal{V}}$} is a regular distribution.
\end{Definition}
Note that \smash{$\vprol{\mathcal{V}}$} is tangent to $\bar E$ seen as a submanifold in $J^{\infty}_VE$. Moreover, the restriction of \smash{$\vprol{\mathcal{V}}$} to~$\bar E$ remains regular if~\smash{$\vprol{\mathcal{V}}$} is regular.

\subsection{Weak gauge PDEs}

Now we are ready to introduce our main objects which are gPDEs with the condition $Q^2=0$ relaxed in a controllable way.
\begin{Definition}\label{def:WGPDE}
\quad
\begin{enumerate}\itemsep=0pt
 \item A weak gPDE $(E,Q,\cK,T[1]X)$ is a $\mathbb{Z}$-graded bundle $\pi\colon E\rightarrow T[1]X$ equipped with a~vector field $Q$ such that $\gh Q=1$, $Q\circ\pi^*=\pi^*\circ \dx$ and an involutive vertical distribution~$\mathcal{K}$ satisfying
 \begin{itemize}\itemsep=0pt
 \item $\mathcal{K}$ is compatible with a $Q$-structure on $E$, i.e., $L_Q \mathcal{K}\subseteq\mathcal{K}$,
 \item $[Q,Q]=2Q^2\in \mathcal{K}$,
 \item $\mathcal{K}$ is quasi-regular.
 \end{itemize}
\item A solution of $(E,Q,\cK,T[1]X)$ is a section $\sigma\colon T[1]X\rightarrow E$ if the degree 1 vector field along~$\sigma$ defined as $\mathcal{R}_{\sigma}:=\dx\circ \sigma^*-\sigma^*\circ Q$, satisfies
\begin{equation}\label{weom}
\quad \mathcal{R}_{\sigma}\in\sigma^*\mathcal{K},
\end{equation}
where $\sigma^*\mathcal{K}$ denotes the pullback of the distribution $\mathcal{K}$ by $\sigma$.
\item Infinitesimal gauge transformations of a given section $\sigma$ are defined by
\begin{equation}\label{wgc}
 \delta\sigma^*=\sigma^*\circ [Q,Y],
\end{equation}
 where $Y$ is a vertical vector field of degree $-1$ and $L_Y\mathcal{K}\subseteq\mathcal{K}$. In a similar way, one defines gauge for gauge symmetries.
 \item Two solutions differing by an algebraic gauge equivalence generated by{\samepage
 \begin{equation*}
 \delta\sigma^*_{\rm alg}=\sigma^*\circ K,
 \end{equation*}
where $K\in \mathcal{K}$, $\gh{K}=0$, are considered equivalent.}
\end{enumerate}
\end{Definition}
Recall that the distributions involved are assumed to be finitely-generated submodules of vector fields. Let us check that the gauge transformation \eqref{wgc} preserves the form of \eqref{weom}. To this end, let us denote by $\mathcal{A}$ the subalgebra of functions annihilated by $\mathcal{K}$. Proposition \ref{prop2} then implies that for any function $f\in\mathcal{A}$ we have
\begin{align*}
\mathcal{R}_{\sigma+\delta\sigma}(f)&{}=\mathcal{R}_{\sigma}(f)+(\dx\circ\sigma^*\circ[Y,Q]-\sigma^*\circ[Y,Q]\circ Q)(f)\\
&{}=(\dx\circ\sigma^*\circ YQ(f)-\sigma^*\circ Q\circ YQ(f))=0,
\end{align*}
where we used Proposition \ref{prop1}, equations of motion~\ref{weom}, and the fact that $Y$ preserves $\mathcal{K}$.

Let us give an important class of examples of weak gPDEs (wgPDEs in what follows).

\begin{Example} \label{ex:presymp}
 Let $\pi\colon E\rightarrow T[1]X$, $\dim{X}=n$, be a $\mathbb{Z}$-graded fiber bundle equipped with a~2-form $\omega$ of degree $n-1$, a 0-form $\mathcal{L}$ of degree $n$ and a degree 1 vector field $Q$ satisfying $Q\circ\pi^*=\pi^*\circ {\rm d}_X$ and
 \begin{equation*}
 {\rm d}\omega=0,\qquad i_Q\omega+{\rm d}\mathcal{L}\in \mathcal{I},\qquad \frac{1}{2}i_Qi_Q\omega+Q\mathcal{L}=0,
 \end{equation*}
 where $\mathcal{I}$ is an ideal generated by the differential forms \smash{$\pi^*\alpha$ with $\alpha\in \Gamma\bigl(\bigwedge^{k>0}(T[1]X)\bigr)$}.\footnote{\changed{We use $\Gamma\bigl(\bigwedge^{k>0}(T[1]X)\bigr)$ to denote the algebra of differential form of positive form-degree on graded manifold~$T[1]X$. Locally, this algebra is generated by $x^a$, $\theta^a$, ${\rm d}x^a$, ${\rm d}\theta^a$.}} This data defines the so-called presymplectic BV-AKSZ formulation (also known as a weak presymplectic gPDE)~\cite{Dneprov:2024cvt, Grigoriev:2022zlq}, which is known to encode the local BV system describing the underlying gauge theory. As we show below this also defines a wgPDE.

 Consider a vertical distribution $\mathcal{K}$ on $E$ generated by vertical vector fields $K$ satisfying
 \begin{equation}\label{dis1}
 i_K\omega\in\mathcal{I}.
 \end{equation}
 Let $K_1$, $K_2$ be vector fields from $\mathcal{K}$ and $\alpha\in\mathcal{I}$. Acting on \eqref{dis1} with Lie derivative $L_{K_2}$, we have
 \begin{gather*}
 L_{K_2}i_{K_1}\omega=L_{K_2}\alpha,\\
 i_{[K_2,K_1]}\omega\pm i_{K_1}L_{K_2}\omega=i_{K_2}{\rm d}\alpha,\\
 i_{[K_2,K_1]}\omega\pm i_{K_1}{\rm d}i_{K_2}\omega=i_{K_2}{\rm d}\alpha,\\
 i_{[K_2,K_1]}\omega=i_{K_2}{\rm d}\alpha\pm i_{K_1}{\rm d}\alpha^{\prime},
 \end{gather*}
 where the sign $\pm$ depends on the degrees of $K_1$, $K_2$. Since $\alpha$ and $\alpha^{\prime}$ are 1-forms from the ideal~$\mathcal{I}$, it follows that $i_{[K_2,K_1]}\omega\in \mathcal{I}$. Hence, $[K_1,K_2]\in \mathcal{K}$. Acting on \eqref{dis1} with $L_Q$ and taking into account that $Q$ preserves $\mathcal{I}$, we obtain
 \begin{equation*}
 i_{[Q,K]}\omega\in\mathcal{I},
 \end{equation*}
 and hence $[Q,K]\in \mathcal{K}$.

 Let us check the condition $i_{[Q,Q]}\omega\in \mathcal{I}$
 \begin{align}
 i_{[Q,Q]}\omega&{}=(L_Qi_Q-i_QL_Q)\omega=(i_Q{\rm d}-{\rm d}i_Q)i_Q\omega-i_Q(i_Q{\rm d}-{\rm d}i_Q)\omega\nonumber\\
 &{}=i_Q{\rm d}i_Q\omega-{\rm d}i_Qi_Q\omega+i_Q{\rm d}i_Q\omega=2i_Q({\rm d}i_Q\omega)+2{\rm d}(Q\mathcal{L})\nonumber\\
 &{}=2i_Q(i_Q{\rm d}-L_Q)\omega+2{\rm d}(i_Q{\rm d}\mathcal{L})=-2i_QL_Q\omega+2(i_Q{\rm d}-L_Q){\rm d}\mathcal{L}\nonumber\\
 &{}=-2i_QL_Q\omega-2L_Q(\mathcal{I}-i_Q\omega)=2(L_Qi_Q-i_QL_Q)\omega-2L_Q\mathcal{I}.\label{q1}
 \end{align}
Combining left- and right-hand sides in \eqref{q1}, we get
\begin{equation*}
 i_{[Q,Q]}\omega=2L_Q\mathcal{I} ,
\end{equation*}
where $2L_Q\mathcal{I}$ belongs to $\mathcal{I}$ and hence $i_{[Q,Q]}\omega\in \mathcal{I}$. Therefore, the vertical distribution $\mathcal{K}$ is involutive and $Q$-invariant, and $[Q,Q]\in \mathcal{K}$ so that we have indeed arrived at a weak gPDE. This gives a variety of examples of wgPDEs describing Lagrangian gauge systems, see~\cite{Dneprov:2024cvt, Grigoriev:2022zlq} for further details and examples.
 \end{Example}

\subsection{Local BV system from wgPDE}
\label{sec:thm}
\begin{Theorem}\label{prop:main}
Let $(E,Q,\cK,T[1]X)$ be a weak gPDE such that the prolongation \smash{$\vprol{\cK}$} of $\cK$ to vertical jets \smash{$J^\infty_VE$} is a fibration. Then it induces an equivalent local BV system whose underlying~bundle is~\smash{$\bar E/\vprol{\cK}_0 \to X$}, where $\bar E$ is a restriction of $J^\infty_VE$ to $X \subset T[1]X$
and \smash{$\vprol{\cK}_0$} is a restriction of \smash{$\vprol{\cK}$} to $\bar E$. In particular, any weak gPDE defines a local BV system, at least locally.
\end{Theorem}
Before giving a proof it is useful to introduce the notion of a weak local BV system.
\begin{Definition}
 A weak local BV system $(J^{\infty}\cE,s,\cN,X)$ is a graded fiber bundle $\cE\rightarrow X$ endowed with an involutive, vertical, regular distribution $\cN$
 and a degree 1 evolutionary vector field $s$ on $J^{\infty}\cE$, which satisfy the following conditions:
 \begin{enumerate}\itemsep=0pt
 \item The jet-prolongation $\bar\cN$ of $\cN$ is $s $-invariant, i.e., $L_s\bar\cN\subset \bar\cN$.
 \item $[s,s]=2s^2\in \bar\cN$.
 \end{enumerate}
\end{Definition}
One can easily specialise the definition of equations of motion and gauge symmetries to this setup.

The following proposition explains the relation to local BV systems.

\begin{Proposition}\label{ibv}
 Let $(J^{\infty}\mathcal{E},s,\cN,X)$ be a weak local BV system such that $\cN$ defines a fibration. Then the quotient $\mathcal{E}/\cN$ is naturally a local BV-system. In particular, any weak local BV system defines a local BV system, at least locally.
\end{Proposition}
\begin{proof}
By assumption $\cN$ defines a fibration and hence the quotient space $\mathcal{E}/\cN$ is well defined. Furthermore, since $\cN$ is vertical, $\mathcal{E}/\cN$ is naturally a fiber bundle over the same base~$X$.

 Let $\bar\cN\subset TJ^{\infty}\mathcal{E}$ denotes a jet-prolongation of $\cN$. Because $\bar\cN$ is generated by evolutionary vector field, it is preserved by the vector field of the form $\total(h)$, for all $h\in \mathfrak{X}(X)$ and, in particular, by total derivatives \smash{$D_a=\total{\dl{x^a}}$}. It follows that
 vector field of the form $\total(h)$, $h\in \mathfrak{X}(X)$, are well defined on the quotient $J^{\infty}\cE/\bar \cN$, determining the Cartan distribution therein. In fact, $J^{\infty}\cE/\bar \cN$ is isomorphic to $J^\infty(\cE/\cN)$.

 Because BRST differential $s$ preserves $\bar\cN$, it induces a well-defined vector field $s^{\prime}$ on $J^{\infty}(\mathcal{E}/\cN)$. Moreover, $s^2\in \bar\cN$ implies that~$s^\prime$ is nilpotent while $\commut{s}{\total{h}}=0$, for all $h\in \mathfrak{X}(X)$ implies that~$s^\prime$ is evolutionary. Thus we conclude that $J^{\infty}(\mathcal{E}/\cN)$ is endowed with the structure of a~local BV system.
\end{proof}

Now we are ready to give a proof of the main Theorem~\ref{prop:main}.
\begin{proof} Given a weak gPDE $\bigl(E,Q,\mathcal{K},T[1]X\bigr)$, its prolongation \smash{$\bigl(J^{\infty}E,\bar{Q},\bar\cK,T[1]X\bigr)$} is again a~weak gPDE. Here, $\bar{Q}$, $\bar\cK$ are prolongations of $Q$ and ${\cK}$ respectively.
The defining relations~\eqref{def:WGPDE} of wgPDE imply
\begin{equation}\label{pr}
 \bigl[\bar\cK,\bar\cK\bigr]\subseteq \bar\cK, \qquad
 \bigl[\bar{Q},\bar{Q}\bigr]\in \bar\cK,\qquad L_{\bar{Q}}\bar\cK\subseteq \bar\cK.
\end{equation}
Moreover, $\bar\cK$ is vertical.

By definition of prolongation, $\bar Q$ projects to $Q$ by $J^{\infty}(E) \to E$ and its vertical part is evolutionary. It follows $\bar{Q}$ can be decomposed as $\bar{Q}=D+s$, where $D=\total(\dx)$ and $s$ is an evolutionary vector field, see, e.g.,~\cite{Grigoriev:2022zlq, Grigoriev:2019ojp} for more details. Then, the last two conditions in~\eqref{pr} give~us
\begin{gather}
 [s,s] \in \bar\cK,\nonumber\\
 L_D K+L_s K\in \bar\cK, \quad \forall K\in \bar\cK.\label{prss}
\end{gather}
Since $\bar\cK$ is the prolongation of a vertical distribution, $D$ preserves $\bar\cK$ and hence for all $K\in\bar\cK$ one has $[D,K]\in \bar\cK$. The last equation in \eqref{prss} then implies that $\bar\cK$ is $s$-invariant as well.

The space-time manifold $X$ can be considered as a zero section in $T[1]X$. The restriction of~$J^{\infty}E$ to $X\subset T[1]X$ can be identified with $J^\infty\bar E$ for another bundle $\bar E$ over $X$ (see Section~\ref{sec:vert-jets}). Moreover, Cartan distribution on $J^{\infty}\bar E$
is generated by the restrictions to $J^{\infty}\bar E$ of vector field of the form $\total{h}$, $h\in \mathfrak{X}(T[1]X)$ with $h$ tangent to $X\subset T[1]X$. Note also that $J^\infty \bar E$ is naturally a~submanifold in $J^\infty E$.

It turns out that $J^\infty \bar E$ is equipped with all the structures of a weak local BV-system. Indeed, because $s$ is vertical it is automatically tangent to $J^\infty \bar E$ seen as a submanifold in $J^{\infty}E$. Moreover, its restriction to $J^\infty\bar E$ is evolutionary therein because $s$ commutes with the vector field of the form
$\total{h}$, $h\in \mathfrak{X}(T[1]X)$. Distribution $\bar\cK$ is also vertical and hence restricts to $J^{\infty}\bar E$ as well, giving a regular distribution $\bar\cK_0$ generated by evolutionary vector field. Note that $\bar\cK_0$ is regular because it originates from the prolongation of $\cK$ which is quasi-regular by assumption.

Recall, that on $J^\infty E$ we had
$s^2\in \bar\cK$, $L_s \bar\cK \subset \bar\cK$.
It follows the same relations hold for $\bar\cK_0=\bar\cK|_{J^\infty(\bar E)}$ and the restrictions of $s$
to $J^\infty\bar E$. Furthermore, \smash{$\bar\cK_0$}
can be also obtained as follows: (i) taking the vertical prolongation of $\cK$ from $E$ to its vertical jets \smash{$J_V^\infty E$}; (ii) then restricting the resulting distribution \smash{$\vprol{\cK}$} to $\bar E$ seen as \smash{$J_V^\infty E|_X$}, giving a regular vertical distribution~\smash{$\vprol{\cK}_0$} on~\smash{$\bar E$}; (iii) taking the prolongation of \smash{$\vprol{\cK}_0$} to \smash{$J^\infty\bar E$}. In this way, one concludes that
$\bar\cK_0$ is a jet-prolongation to $J^\infty\bar E$ of the distribution~\smash{$\vprol{\cK}_0$} on~$\bar E$. It follows \smash{$\bigl(J^{\infty}\bar E,s,\vprol{\cK}_0,X\bigr)$}, where by some abuse of notations $s$ also denotes its restriction to $J^\infty\bar E$, is a weak local BV system. Its construction from the starting point wgPDE $(E,Q,\cK,T[1]X)$ is canonical.

Finally, Proposition~\ref{ibv} implies that \smash{$\bigl(J^\infty\bar E,s,\bar\cK_0,X\bigr)$}
determines a local BV system whose underlying bundle is \smash{$\bar E/\vprol{\cK}_0$}.
\end{proof}

\subsection{Inequivalent solutions}

As we have just seen a wgPDE defines a local BV system, at least locally. This gives wgPDEs the unambiguous field-theoretical interpretation. In particular, solutions of a wgPDE related by algebraic gauge transformations, as defined in Definition~\ref{def:WGPDE}, correspond to one and the same solution of the underlying local BV system.

It is desirable to define inequivalent solutions of wgPDEs from the very start without resorting to algebraic equivalence, at least in the case where $\cK$ is sufficiently good. To this end, we restrict ourselves to the case where the regular distribution \smash{$\vprol{\cK}$} (the vertical prolongation of $\cK$) defines a~fibration of $J^\infty_VE$. Of course, this is always the case locally, as by assumption $\cK$ is quasi-regular.

To define inequivalent solutions as subbundles, consider vertical jets of sections (as opposed to supersections) of $E$ and denote it by $J^\infty_{V0}E$. Note that \smash{$J^\infty_{V0}E$} is naturally a subbundle in \smash{$J^\infty_{V}E$} and its restriction to $X$ is nothing but the body of $\bar E$. The ghost degree zero component of $\cN$ determines an involutive vertical distribution on \smash{$J^\infty_{V0}E$}. We say that a subbundle \smash{$\Sigma\subset J^\infty_{V0}E$} is an integral subbundle of $J^\infty_{V0}E$ if its fibres are integral submanifolds of the above distribution on~\smash{$J^\infty_{V0}E$}.\looseness=-1

Any section $\sigma\colon T[1]X\to E$ defines a section of $J^\infty_{V0}E$ via vertical prolongation. It is easy to check that two solutions are algebraically equivalent if their prolongations to $J^\infty_{V0} E$ belong to the same integral subbundle of $J^\infty_{V0} E$. In other words, an equivalence class of solutions to $(E,Q,\cK,T[1]X)$ is an integral subbundle of $J^\infty_{V0} E$ such that there exists a solution $\sigma\colon T[1]X \to E$ such that its prolongation to $J^\infty_{V0} E$ belongs to this integral subbundle.

\section{Examples} \label{sec:examples}

\subsection{Scalar field }
We begin with the simplest example of a local theory without gauge freedom, namely, a free massive scalar field in Minkowski spacetime. In this case the fiber bundle underlying the minimal wgPDE formulation is $E\rightarrow T[1]X$, where $X=\mathbb{R}^{1,3}$ is a four-dimensional Minkowski space with the metric $\eta_{ab}$. The fiber is $\fR^1 \times \fR^{1,3}$ with coordinates $\phi$, $\phi_a$ of ghost degree zero. The adapted local coordinates on $E$ are $(x^a,\theta^a,\phi,\phi^a)$. The $Q$-structure is defined by
\begin{alignat*}{3}
 &Q(x^a)=\theta^a, \qquad&& Q(\theta^a)=0,&\\
 &Q(\phi)=\theta^a\phi_a, \qquad&& Q(\phi^a)=\frac{1}{4}m^2\theta^a\phi,&
\end{alignat*}
where $m\in\mathbb{R}$ is the mass. Note that \smash{$Q^2(\phi^a)=-\frac{1}{4}m^2\theta^a\theta^b\phi_b$} and hence $Q$ is not nilpotent in general. The generalization to the case of interacting field and curved spacetime is straightforward.

Let us consider the distribution $\mathcal{K}$ generated by
\begin{gather}
 K^{(1)}_{ab}:=\theta_a\frac{\partial}{\partial \phi^b}-\frac{1}{4}\eta_{ab}\theta^c\frac{\partial}{\partial\phi^c},\qquad K^{(2)}_{ab}:=\theta_a\theta_b\frac{\partial}{\partial\phi},\qquad
 K^{(3)}_{abc}:=\theta_a\theta_b\frac{\partial}{\partial\phi^c}.\label{scald}
 \end{gather}
It is easy to verify that
\begin{gather*}
 \bigl[Q, K^{(1)}_{ab}\bigr]=K^{(2)}_{ab},\qquad
 \bigl[Q, K^{(2)}_{ab}\bigr]=m^2\theta_a\theta^cK^{(1)}_{cb},\qquad
 \bigl[Q, K^{(3)}_{abc}\bigr]=-\theta_cK^{(2)}_{ab}
\end{gather*}
as well as \smash{$Q^2=-m^2\theta^a\phi^b K^{(1)}_{ab}$}.
Thus, $Q$ preserves $\cK$ and all the conditions of Definition~\ref{def:WGPDE} are fulfilled. Of course, this system is Lagrangian and the above distribution can be defined as the kernel distribution of the presymplectic structure $\omega={\rm d}\bigl(\epsilon_{abcd}\theta^a\theta^b\theta^c \phi^d {\rm d}\phi\bigr)$ of the minimal presymplectic BV-AKSZ formulation of the scalar field.

Let us now define the prolongation of the distribution \eqref{scald} to $\bar{E}$.
Specializing to the case at~hand the coordinate system on $J^\infty_VE$ introduced in Section~\ref{sec:vert-jets}, we arrive at coordinates
$x^a$, $\theta^a$, $\bar \phi_{|\dots}$, $\bar \phi^a_{|\dots}$ on $J^\infty_VE$
such that \smash{$D^{(\theta)}_a=\dl{\theta^a}$} and the projection $\pi^\infty_V\colon J^\infty_VE \to E$ acts as
\begin{gather*}
 (\pi^\infty_V)^*(\phi)=\bar\phi+\theta^a\bar\phi_{|a}+\frac{1}{2}\theta^a\theta^b\bar\phi_{|ab}+\cdots,\\
 (\pi^\infty_V)^*(\phi^a)=\bar\phi^a+\theta^b\bar\phi^a_{\,\,|b}+\frac{1}{2}\theta^b\theta^c\bar\phi^a_{\,\,|bc}+\cdots,
 \end{gather*}
where \smash{${\rm gh}\bigl(\bar\phi_{|a_1\dots a_k}\bigr)={\rm gh}\bigl(\bar\phi^a_{\,\,|a_1\dots a_k}\bigr)=-k$}.

If \smash{\smash{$\vprol{\cK}$}} denotes the prolongation of $\cK$ to $J^\infty_VE$, then the restriction of \smash{$\vprol{\cK}$} to
$\bar E\subset J^\infty_VE$ is denoted by \smash{$\vprol{\cK}_0$} and is generated by the following vector fields
\begin{gather*}
 \biggl(\frac{1}{4}\eta_{ab}\delta^m_n-\eta_{an}\delta^m_b\biggr)\frac{\partial}{\partial \bar{\phi}^m_{\,\,\,\,|n}},
 \qquad \frac{\partial}{\partial \bar\phi^a_{|bc}},
 \qquad \frac{\partial}{\partial \bar\phi^a_{|bcd}},
 \qquad \frac{\partial}{\partial \bar\phi^a_{|bcde}},\\
 \frac{\partial}{\partial\bar{\phi}_{|ab}},\qquad
 \frac{\partial}{\partial\bar{\phi}_{|abc}}, \qquad
 \frac{\partial}{\partial\bar{\phi}_{|abcd}}.
 \end{gather*}
It is easy to check that the subalgebra of functions on $\bar E$, annihilated by \smash{$\vprol{\cK}_0$} is generated by $x^a$ and
\begin{gather}
 \bar\phi,\qquad \bar\phi^a,\qquad
 \bar\phi^*\equiv\bar\phi^a_{\,\,|a},\qquad \bar\phi^*_a\equiv\bar\phi_{|a},\label{scaldalg}
 \end{gather}
giving the field content of the standard BV formulation of the first-order form of the scalar field.

We have seen in Section~\ref{sec:thm} that the BRST differential $s$ of the underlying local BV system is the vertical part of the prolongation of $Q$ to $J^\infty E$, restricted to $\bar E$. In the case at hand, it is easy to find how $s$ acts on the coordinates \eqref{scaldalg},
\begin{alignat*}{3}
 &s\bigl(\bar\phi\bigr)=0,\qquad&& s\bigl(\bar\phi^a\bigr)=0,&\\
 &s\bigl(\bar\phi^*\bigr)=D_a\bar\phi^a-m^2\bar\phi,\qquad&& s\bigl(\bar\phi^*_a\bigr)=D_a\bar\phi-\eta_{ab}\bar\phi^b.&
 \end{alignat*}
This is of course the standard BRST-differential for the BV extension of the first-order formulation of the scalar field.

\subsection{Self-Dual Yang--Mills}
Let us take as $X$ the Euclidean space $\mathbb{R}^4$ with the standard metric $\delta_{ab}$ and consider a wgPDE $E\rightarrow T[1]\mathbb{R}^4$, with the fiber being $\algg[1]$, where $\algg$ is a real Lie algebra. The local coordinates on $E$ are $\bigl(x^a,\theta^a,C^A\bigr)$. Instead of $C^A$, it is convenient to work in terms of $\algg$-valued coordinate function $C=C^At_A$, where $t_A$ denote a basis in $\algg$. The $Q$-structure is then defined as
\begin{gather*}
 Q(x^a)=\theta^a,\qquad Q(\theta^a)=0,\qquad
 Q(C)=-\frac{1}{2}\lb C,C\rb,
\end{gather*}
where $\lb\cdot,\cdot\rb$ denotes the Lie bracket in $\algg$. Note that the above data defines a gPDE which describe a zero-curvature equation for a $\algg$-valued 1-form. In other words, if we take $\cK=0$, we are dealing with a topological system.

To describe the self-dual YM which is not topological, we take $\cK$ to be generated by
\begin{equation*}
 K^{ab}_{A}=\biggl(\theta^a\theta^b+\frac12\epsilon^{ab}{}_{cd}\theta^c\theta^d\biggr)\frac{\partial}{\partial C^A},
\end{equation*}
where $\epsilon^{ab}{}_{cd}=\delta_{cm}\delta_{dn}\epsilon^{abmn}$.

It is easy to check that
\begin{equation*}
 \bigl[Q,K^{ab}_{A}\bigr]=f_{AB}{}^{C}C^B K^{ab}_{C},
\end{equation*}
where $f_{AB}{}^{C}$ are the structure constants: $\lb t_A,t_B \rb = f_{AB} {}^{C}t_C$.
In this case, $Q$ is nilpotent due to the graded Jacobi identities. Thus, all axioms of Definition~\ref{def:WGPDE} are satisfied.

Now let us discuss the BV-system induced by the above wgPDE. Coordinate system on $J^{\infty}_VE$ is given by $\bigl(x^a,\theta^a,\bar{C}_{|\dots}\bigr)$, where ${\rm gh}\bigl(\bar{C}_{|a_1\dots a_k}\bigr)=1-k$, and the projection $\pi^{\infty}_{V}\colon J^{\infty}_V E\rightarrow E$ acts in a standard way as in \eqref{projection-vert-jets}.

The distribution \smash{$\vprol{\cK}_0$} on $\bar E\subset J^\infty_VE$ induced by $\cK$ is generated by the following vector fields:
\begin{gather*}
 \frac{\partial}{\partial \bar{C}^A_{|ab}}+\frac{1}{2}\epsilon^{ab}_{\,\,\,\,\,\,mn}\frac{\partial}{\partial \bar{C}^A_{|mn}},
 \qquad \frac{\partial}{\partial \bar{C}^A_{|abc}},
 \qquad \frac{\partial}{\partial \bar{C}^A_{|abcd}},
\end{gather*}
where the second and the third groups are the prolongations of $\theta^aK^{bc}_A$ and $\theta^a\theta^bK^{cd}_A$, respectively.

It is easy to check that the subalgebra of functions on $\bar{E}$ annihilated by \smash{$\vprol{\cK}_0$} is generated by~$x^a$ and
\begin{gather*}
 \bar{C},\qquad A_a\equiv \bar{C}_{|a},\qquad
 \mathcal{F}^{*-}_{ab}\equiv \bar{C}_{|ab}-\frac{1}{2}\epsilon_{ab}{}^{cd}\bar{C}_{|cd}.\label{minYM}
 \end{gather*}

The induced BRST differential $s$ acts on \eqref{minYM} as follows:
\begin{gather*}
 s( \mathcal{F}^{*-}_{ab})=-(D_aA_b-D_bA_a+\lb A_a,A_b\rb)^{-}-\lb\mathcal{F}^{*-}_{ab},\bar{C}\rb,\\
 s(\bar{C})=-\frac{1}{2}\lb\bar{C},\bar{C}\rb,\quad s(A_a)=\partial_a\bar{C}+\lb A_a,\bar{C}\rb,
\end{gather*}
where $D_a$ is the total derivative on \smash{$J^\infty \bigl(\bar E/\vprol{\cK}_0\bigr)$}. This defines the BRST differential of the local BV system on \smash{$J^\infty \bigl(\bar E/\vprol{\cK}_0\bigr)$}. It is easy to see that this is indeed the BRST differential of the self-dual Yang--Mills theory. Note that the BRST complex for a version of this system was discussed in~\cite{Lyakhovich:2007cw}.

Let us finally comment on the relation between weak gPDEs and the AKSZ sigma models, where the algebra $\cC^\infty(T[1]X)$ of functions on the source (this is of course just the exterior algebra of $X$) is replace by a more general differential graded commutative algebra, for instance, not freely generated one. Such AKSZ-like formulations were put-forward in~\cite{costello2011renormalization}, see also~\cite{Bonechi:2022aji,Bonechi:2009kx,Grigoriev:2025vsl}.

Without trying to be general, let us restrict ourselves to the above example of SDYM. In this case, it is easy to observe that taking the quotient by $\cK$ prolonged to the space of super-sections, simply amounts to keeping only certain components of the differential forms on $X$. Because distribution $\cK$ has the factorized form \big(there is no mixing between $x$, $\theta$ and $C^A$\big), taking the quotient only affects the base space component of the differential forms and the remaining fields can be understood as those parametrizing supermaps from the certain quotient of $\cC^\infty(T[1]X)$ to the fibre $\algg[1]$. This can be made more precise with the following proposition.

\begin{Proposition}
Suppose that $(E,Q,T[1]X,\cK)$ be a wgPDE of AKSZ type, that is, $E=T[1]X\times F$, $Q=\dx+q$. Moreover, let $\cK$ be of the factorized form, i.e., generated by vector fields of the form
$h_\alpha\dl{C^A}$, where $h_\alpha\in \cC^\infty(T[1]X)$ and $C^A$ are local coordinates on $F$. In addition, we assume that $\gh{h_\alpha}>0$ so that $h_\alpha$ do not restrict the base $X$. Then the quotient of $\mathrm{Smaps}(T[1]X,F)$ by the prolongation of $\cK$ is isomorphic $($as $Q$-manifolds$)$ to the space of super-homomorphisms from $\cC^\infty(F)$ to $\cC^\infty(T[1]X)/I(h_\alpha)$, where $I(h_\alpha)$ is the ideal in
$\cC^\infty(T[1]X)$ generated by $h_\alpha$.\footnote{Alternatively,
one can speak of supermaps from $\mathrm{Spec}(\cC^\infty(T[1]X)/I(h_\alpha))$ to $F$, where $\mathrm{Spec}(\algA)$ is the ``space'' whose algebra of functions is $\algA$.}
\end{Proposition}
\begin{proof}
Because $h_\alpha$ do not restrict $X$ it is enough to check the statement at a given point $x\in X$, i.e., to replace $C^\infty(T[1]X)$ with a finite dimensional algebra $C^\infty(T_x[1]X)$. Without loss of generality one can assume that $e_\alpha=h_\alpha|_{T_x[1]X}$ form a basis in $I(h_\alpha)|_{T_x[1]X}$. Let $e_i \in C^\infty(T_x[1]X)$ be such that $\{e_i,e_\alpha\}$ form a basis in $C^\infty(T_x[1]X)$. Then coordinates of the space of supermaps can be introduced as
\begin{equation}
{\rm ev}^*\bigl(C^A\bigr)=\psi^{iA}e_i+\psi^{\alpha A} e_\alpha,
\end{equation}
where ${\rm ev}^*$ is the pullback by the evaluation map ${\rm ev}\colon {\rm Smaps}(T_x[1]X,F) \times T_x[1]X \to F$, see, e.g.,~\cite{Roytenberg:2006qz} for details on spaces of supermaps and map $ev$. In these coordinates the prolongation of \smash{$K_{\alpha A}=e_\alpha \dl{C^A}$} to the space of supermaps reads as
\begin{equation*}
{\bar K}_{\alpha A}=\dl{\psi^{\alpha A}}.
\end{equation*}
Functions on the quotient by $\bar \cK$
are clearly the $\psi^{\alpha A}$-independent ones and they can be identified with the functions on the space of superhomomorphsims from $\cC^\infty(F)$ to $\cC^\infty(T_x[1]X)/I(e_\alpha)$. Furthermore, under this identification the $Q$-structure induced by $\dx+q$ on the space of super-homomorphisms coincides with the usual wgPDE one on the quotient by the prolonged $\cK$ is easily seen by direct computation. Note that $\dx$ preserves $I(h_\alpha)$ and hence descends to the quotient because $\cK$ is by definition $Q$-invariant and $Q$ is a product $Q$-structure, i.e., $Q=\dx+q$.
\end{proof}

Applying the above statement to wgPDE describing the self-dual YM equations, we find that the resulting system is the AKSZ sigma model with the same target and source being the quotient of $T[1]\fR^4$ by the ideal generated by $\theta^a\theta^b+\half\epsilon^{ab}{}_{cd}\theta^c\theta^d$. This gives a derivation of the AKSZ formulation of self-dual YM equations in terms of non-freely generated source space DGCA similar to the initial proposal~\cite{costello2011renormalization} by K.~Costello.

\subsection{Finite jets as a weak gPDE}
Let $\mathcal{E}\rightarrow X$ be a fibre bundle. We show that its associated finite jet bundle $J^k\cE$ naturally gives rise to a weak gPDE $\bigl(E^k,Q,\cK,T[1]X\bigr)$ such that its underlying local BV system is equivalent to $J^\infty\cE$ equipped with the trivial BRST differential. In particular, inequivalent solutions of $E^k$
are one-to-one with sections of $\cE$.

Although $\bigl(E^k,Q,\cK,T[1]X\bigr)$ can be defined in a coordinate-free way, we present the construction in local coordinates to simplify the exposition and make it more explicit. Let $(x^a,u)$ be local coordinates on $\cE$ such that $x^a$ are coordinates on $X$ pulled back to $\cE$ and $u$ is a~fibre coordinate. By slight abuse of notations, the induced coordinates on $J^k\mathcal{E}$ are denoted by~$(x^a,u,u_a,\dots,u_{a_1\dots a_k})$. The Cartan distribution $\mathcal{C}$ on $J^k\mathcal{E}$ is generated by the following vector fields:
\begin{gather}\label{cartan}
 \mathcal{D}_a=\frac{\partial}{\partial x^a}+\sum^{k-1}_{i=0}u_{a b_1\dots b_i}\frac{\partial}{\partial u_{b_1\dots b_{i}}}, \qquad V^{a_1\dots a_k}=\frac{\partial}{\partial u_{a_1\dots a_k}},
\end{gather}
where $\mathcal{D}_a$ is often refereed to as the truncated total derivative. Note that in contrast to the Cartan distribution on infinite jet-bundles this one is not involutive, see, e.g.,~\cite{Bocharov:1999}.

Now we turn to the construction of the associated wgPDE. First we take $E^m \to T[1]X$, $m\leq k$ to be the pullback of $J^m\mathcal{E}$ by the projection $p\colon T[1]X\rightarrow X$. \changed{The natural coordinates on the total space of $E^m\to T[1]X$} are then $(x^a,\theta^a,u,u_a,\dots,u_{a_1\dots a_m})$. Note that there are natural projections $E^{m}\to E^{m-1}$ induced by the canonical projections $J^m\cE\to J^{m-1}\cE $. The algebra $C^{\infty}\bigl(E^k\bigr)$ of functions on $E^k$ has the following natural filtration:
\begin{equation*}
\cF_0\equiv C^{\infty}(\cE) \subset \cF_1\subset \dots \subset \cF_{k-1}\subset \cF_k\equiv C^{\infty}\bigl(E^k\bigr),
\end{equation*}
where $\cF_m$ is the subalgebra of functions pulled back from $E^m$, i.e., functions depending on jets of order not higher than $m$. The following derivation $Q\colon \cF_{k-1}\to \cF_{k}$ is well defined:
\begin{equation*}
Q=\theta^a \cD_a,
\end{equation*}
where $\cD_a$ is the truncated total derivative from~\eqref{cartan}. This can be easily checked by performing the coordinate change on $\cE$ and the associated coordinate change on $E^k$. Note that the above~$Q$ can be formally applied to functions from $\cF_{k}$ but the result is coordinate-dependent.

Furthermore, on $E^k$ one can define the distribution $\cK$ generated by the following vector fields:
\begin{gather}\label{closure}
 \theta^{b_1}\cdots\theta^{b_r}\frac{\partial}{\partial u_{a_1\dots a_m}}, \qquad r+m=k+1, \ r\geq 1 ,\ m\geq 0.
\end{gather}
Note that these vector fields clearly vanish for $r>n\equiv \dim(X)$ and $m<k-n+1$ and $m>k$. Note that although generators \eqref{closure} are generally coordinate-dependent the distribution itself is well defined.

Although $Q$ is not invariantly defined on $\cF_k\equiv \cC^\infty\bigl(E^k\bigr)$, in each coordinate patch it defines
a unique equivalence class of locally-defined vector fields modulo $\cK$ and the equivalence classes agree on the overlap of patches. In this sense the equivalence class of $Q$ is well-defined globally. Modulo this subtlety $\bigl(E^k,Q,\cK,T[1]X\bigr)$ is a weak gPDE. Indeed, it is easy to check that all the axioms are satisfied: $\cK$ is $Q$-invariant and $Q^2\in \cK$.

Let us see what systems does the constructed wgPDE determine. To this end we construct its associated local BV system. According to the Theorem~\ref{prop:main}, the fiber-bundle underlying the local BV system is the quotient of $\bar E$ by the distribution \smash{$\vprol{\cK}_0$} which is the restriction to $\bar E \subset J^{\infty}_VE$ of the prolongation of $\mathcal{K}$ to $J^{\infty}_VE$. In our case, this distribution is generated by the following vector fields:
\begin{equation}\label{prolongd}
 \begin{gathered}
\frac{\partial}{\partial\bar u_{a_1\dots a_{m}|b_1\dots b_{r}}} ,\qquad m+r\geq k+1,
 \end{gathered}
\end{equation}
\changed{where $\bar{u}_{a_1\dots a_m|b_1\dots b_r}$ denote the fibre coordinates on $\bar{E}$ induced by $u_{a_1\dots a_m}$, see Section~\ref{sec:vert-jets}.} It~follows that the subalgebra of functions in the kernel of \smash{$\vprol{\cK}_0$} is generated by $x^a$ and{\samepage
\begin{gather*}
 \bar{u}_{a_1\dots a_m|b_1\dots b_r},
 \qquad m+r\leq k.
 \end{gather*}
In other words, all the coordinates with the total number of indices $\leq k$ are in the kernel of~\eqref{prolongd}.}

Now we turn to the BRST differential induced by $Q$ on $J^\infty\bigl(\bar E/\vprol{\cK}_0\bigr)$. It has the following structure (see, e.g.,~\cite{Barnich:2010sw,Grigoriev:2019ojp}):
$s={\rm d}^F-\sigma^F$,
where ${\rm d}^F$ arises from a vertical part of the prolongation of \smash{$\dx=\theta^a\dl{x^a}$} and $\sigma^F$ originates from the prolongation of the vector field \smash{$\sigma\equiv Q-\dx=\theta^a\bigl(\cD_a-\dl{x^a}\bigr)$}. Of course, the decomposition of~$Q$ into~$\dx$ and~$\sigma$ is not invariant and hence depends on the trivialisation and is defined only locally. Working locally and using our adapted coordinate system, one observes that $\sigma$ is a vertical vector field on $E$ and hence its prolongation to $J^\infty_VE$ is well defined. Restricting the prolongation to $\bar E$, one gets a vector field on $\bar E$, which preserves \smash{$\vprol{\cK}_0$} and hence defines a vector field $\tilde\sigma_0$ on \smash{$\bar E/\vprol{\cK}_0$}. Explicitly, the action of~$\tilde\sigma_0$ on coordinates on \smash{$\bar E/\vprol{\cK}_0$} is given by
\begin{gather*}
 \tilde\sigma_0 \bar{u}_{a_1\dots a_m|b_1\dots b_r}=(-1)^r\bar{u}_{a_1\dots a_m[b_1|b_2\dots b_r]}, \qquad m+r\leq k.
\end{gather*}
For instance, $\tilde\sigma_0\bar{u}_{|a}=-\bar{u}_{a|}$, \smash{$\tilde\sigma_0\bar{u}_{|ab}=\frac{1}{2}(\bar{u}_{a|b}-\bar{u}_{b|a})$}.

Now we are ready to show that the local BV system \smash{$\bigl(J^\infty\bigl(\bar E/\vprol{\cK}_0\bigr),{\rm d}^F-\sigma^F,X\bigr)$} is equivalent to $J^\infty\cE$ equipped with the trivial BRST differential. To this end, we introduce the auxiliary degree
$\deg$ in \smash{$\cC^\infty\bigl(E^k\bigr)$} according to $\deg(\bar{u}_{a_1\dots a_m})=m$. Then it is easy to check that $\deg \sigma^F=1$ while \smash{$\deg\bigl({\rm d}^F\bigr)=0$}. It follows that if $w^\alpha$, $v^\alpha$ are
coordinates on \smash{$\bar E/\vprol{\cK}_0$} such that $\tilde\sigma_0 w^\alpha=v^\alpha$ these coordinates give rise to generalised auxiliary fields of the local BV system. Indeed, the equation $sw^\alpha=0$ can be uniquely solved with respect to $v^\alpha$ and hence the system can be equivalently reduced to the subbundle of \smash{$J^\infty\bigl(\bar E/\vprol{\cK}_0\bigr)$} singled out by the prolongations of $w^\alpha=0$, $sw^\alpha=0$. Moreover, the resulting subbundle is again a jet bundle. See~\cite{Barnich:2004cr, Grigoriev:2019ojp} for more details on the equivalent reductions.

In the case at hand, it is easy to see that all the coordinates $\bar{u}_{a_1\dots a_m|b_1\dots b_r}$ save for $\bar{u}$ can be split into two subsets $w^\alpha$ and $v^\alpha$ such that $\tilde\sigma_0 w^\alpha=v^\alpha$. To see this, one can package $\bar{u}_{a_1\dots a_m|b_1\dots b_r}$ into the generating function
\begin{equation}\label{genfunc}
 U=\sum_{m,r\geq0}\frac{1}{m!r!}y^{a_1}\cdots y^{a_m}\xi^{b_1}\cdots\xi^{b_r}\bar{u}_{a_1\dots a_m|b_1\dots b_r}\equiv y^{(a)}\xi^{[b]}\bar{u}_{(a)|[b]},
\end{equation}
where $y^a$, $\gh{y^a}=0$ are commuting variables and $\xi^a$, $\gh{\xi^a}=0$ are anti-commuting variables. Then one observes that \smash{$\tilde\sigma_0 U=\xi^a\frac{\partial}{\partial y^a} U$}.
Thanks to the Poincar\'{e} lemma, \smash{$\xi^a\frac{\partial}{\partial y^a}$} does not have cohomology in the space of polynomials in $y^a$, $\xi^a$ except for constants. The same is true for polynomials of total order not higher than~$k$. It follows, there exist homogeneous monomials~$f_\alpha$,~$g_\alpha$ in $y$, $\xi$ such that, together with~1, they form a basis in the space of polynomials in $y$, $\xi$ and~\smash{$\xi^a\frac{\partial}{\partial y^a} f_{\alpha}=g_{\alpha}$}, \smash{$\xi^a\frac{\partial}{\partial y^a} 1=0$}. We then have the following decomposition of \eqref{genfunc}:
\begin{equation*}
 U=\bar{u}+f_{\alpha}v^{\alpha}+g_{\alpha}w^{\alpha},
\end{equation*}
 and thus, coordinates $\bar{u}_{a_1\dots a_m|b_1\dots b_r}$ with $m+r>0$ can be replaced with coordinates $v^{\alpha}$, $w^{\alpha}$ such that $\tilde\sigma_0 w^{\alpha}=v^{\alpha}$.

Finally, the elimination of $w^\alpha$, $v^\alpha$ leaves us with just coordinates $\bar{u}$, i.e., the fibre bundle
underlying the reduced local BV system is the starting point $\cE$. It is clear that the reduced BRST differential vanishes simply by the degree reasoning. Of course, this can also be checked directly.\looseness=-1

\section{Conclusions}

As a concluding remark, let us discuss how weak gauge PDEs arise from
genuine gauge PDEs because this creates an interesting approach to characterise those gauge PDEs that correspond to local gauge theories. More precisely, let $(E,Q,T[1]X)$ be a gPDE equipped with an involutive $Q$-invariant quasi-regular distribution $\cK$ so that $(E,Q,\cK,T[1]X)$ can be also considered as a weak gPDE. This means that $(E,Q,\cK,T[1]X)$ gives rise to two apparently different gauge systems: (i) the one encoded in $(E,Q,T[1]X)$ seen as a gPDE; (ii) the one encoded in $(E,Q,\cK,T[1]X)$ seen as a wgPDE. In particular, the difference between the two is that section $\sigma\colon T[1]X\to E$ is a solution to the gPDE if $\dx\circ \sigma^*-\sigma^* \circ Q=0$ while the condition to be a solution to the wgPDE is much weaker: $\dx\circ \sigma^*-\sigma^* \circ Q \in \sigma^* \circ \cK$.
However, solutions to the wgPDE related by algebraic gauge transformations generated by $\cK$ are to be considered equivalent so that solutions to the gPDE can be one to one with equivalence classes of solutions to the wgPDE. We~say that $\cK$ is complete if the local BV systems determined by $(E,Q,T[1]X)$ and $(E,Q,\cK,T[1]X)$ are equivalent. A variety of examples of complete $\cK$ arise as kernel distributions of complete presymplectic structures on gPDEs describing Lagrangian systems, see~\cite{Dneprov:2022jyn,Dneprov:2024cvt, Grigoriev:2016wmk,Grigoriev:2020xec}.

If the underlying system is local, the corank of $\cK$ (or better the corank of the prolongation of $\cK$ to $J^\infty_V E$ as only this prolongation is assumed regular) is finite-dimensional. Of course, if the number of dependent variables of the underlying PDE is infinite, one needs to replace this with the appropriately modified requirement (e.g., being locally finite dimensional). In the case where $\cK$ arises from a presymplectic structure, as~in~Example~\ref{ex:presymp}, or from a projector, as in Example~\ref{ex:projector}, the condition is that the rank of the presymplectic structure or the projector prolonged to $J^\infty_VE$ is finite.

This observation gives a way to characterise those gauge PDEs that describe genuine local theories as those admitting complete distributions $\cK$ of finite-dimensional corank. The problem of characterising gPDEs describing local systems is well known, for instance, in the context of higher spin gauge theories on AdS space~\cite{Vasiliev:1999ba, Vasiliev:2003ev}, in which case the theory is usually defined from the outset in terms of a free differential algebra which is a particular example of gPDE, see, e.g.,~\cite{Boulanger:2015ova,Didenko:2019xzz,Gelfond:2019tac,Skvortsov:2015lja}.

\appendix
\section{Notations and conventions}
Let $\mathcal{M}$ be a graded manifold and $C^{\infty}(\mathcal{M})$ denote its algebra of functions. A $\mathbb{Z}$-grading on $\mathcal{M}$ corresponds to the ghost degree that appears in the text. The Grassmann parity is denoted by~$\p{A}$ and is given by $\p{A}=\gh{A}\mod 2$ if no physical fermions are present. We use the Koszul sign convention, i.e., for all $f,g\in C^{\infty}(\mathcal{M})$ $f\cdot g=(-1)^{|f||g|}g\cdot f$. Let $\mathfrak{X}(\mathcal{M})$ denote the space of vector fields on $\mathcal{M}$. Vector fields on $\mathcal{M}$ act as left derivations, i.e., for all $V\in \mathfrak{X}(\mathcal{M})$ and $f,g\in C^{\infty}(\mathcal{M})$, the graded Leibniz rule holds $V(fg)=V(f)g+(-1)^{|V||f|}fV(g)$.

The de Rham differential ${\rm d}$ and the interior product $i_V$ can both be seen as vector field on the shifted tangent bundle $T[1]\mathcal{M}$. Let $\bigl(\psi^A,{\rm d}\psi^A\bigr)$ be local coordinates on $T[1]\mathcal{M}$, where $\psi^A$ are base coordinates and ${\rm d}\psi^A$ are fibre ones, $\p{{\rm d}\psi^A}=\p{\psi^A}+1$, then ${\rm d}$ and $i_V$ are given by
\begin{equation*}
 {\rm d}f={\rm d}\psi^A\frac{\partial f}{\partial \psi^A},\qquad i_V{\rm d}f=V(f)=V^A(\psi)\frac{\partial f}{\partial \psi^A}.
\end{equation*}

The components of a differential $n$-form $\alpha$ are introduced as
\begin{equation*}
 \alpha=\frac{1}{n!}{\rm d}\psi^{A_1}\cdots {\rm d}\psi^{A_n}\alpha_{A_1\dots A_n}(\psi).
\end{equation*}
The Lie derivative is defined by the graded version of Cartan's magic formula
\begin{equation*}
 L_V=[i_V,{\rm d}]=i_V{\rm d}+(-1)^{|V|}{\rm d}i_V.
\end{equation*}
Since $|{\rm d}|=1$ and $|i_V|=|V|-1$, the Lie derivative is of degree $|V|$. The following formula is useful for us:{\samepage
\begin{equation*}
 i_{[V,W]}=[L_{V},i_{W}]=L_{V}i_{W}-(-1)^{|V|(|W|-1)}i_{W}L_{V},
\end{equation*}
where $V,W\in \mathfrak{X}(\mathcal{M})$.}

\subsection*{Acknowledgements}
We wish to thank K.~Druzhkov, I.~Krasil'shchik, A.~Mamekin, M.~Markov, A.~Verbovetsky, and especially I.~Dneprov and A.~Kotov for fruitful discussions. The authors are also grateful to the anonymous referees for their valuable suggestions and comments that helped to improve the manuscript.
Maxim Grigoriev supported by the ULYSSE Incentive
Grant for Mobility in Scientific Research [MISU] F.6003.24, F.R.S.-FNRS, Belgium, also at Lebedev Physical Institute and Institute for Theoretical and Mathematical Physics, Lomonosov MSU.


\begin{thebibliography}{99}
\footnotesize\itemsep=-0.5pt

\bibitem{Alexandrov:1995kv}
Alexandrov M., Schwarz A., Zaboronsky O., Kontsevich M., The geometry of the
 master equation and topological quantum field theory,
 \href{https://doi.org/10.1142/S0217751X97001031}{\textit{Internat.~J.~Modern
 Phys.~A}} \textbf{12} (1997), 1405--1429,
 \href{http://arxiv.org/abs/hep-th/9502010}{arXiv:hep-th/9502010}.

\bibitem{Alkalaev:2013hta}
Alkalaev K.B., Grigoriev M., Frame-like {L}agrangians and presymplectic
 {AKSZ}-type sigma model,
 \href{http://doi.org/10.1142/S0217751X14501036}{\textit{Int.~J. Mod.
 Phys.~A}} \textbf{29} (2014), 1450103, 33~pages,
 \href{http://arxiv.org/abs/1312.5296}{arXiv:1312.5296}.

\bibitem{Andersonbook}
Anderson I., The variational bicomplex, {T}echnical Report, Department of
 Mathematics, Utah State University, 1989.

\bibitem{Barnich:1995ap}
Barnich G., Brandt F., Henneaux M., Local {BRST} cohomology in
 {E}instein--{Y}ang--{M}ills theory,
 \href{https://doi.org/10.1016/0550-3213(95)00471-4}{\textit{Nuclear Phys.~B}}
 \textbf{455} (1995), 357--408,
 \href{http://arxiv.org/abs/hep-th/9505173}{arXiv:hep-th/9505173}.

\bibitem{Barnich:1995db}
Barnich G., Brandt F., Henneaux M., Local {BRST} cohomology in the antifield
 formalism.~{I}. {G}eneral theorems,
 \href{https://doi.org/10.1007/BF02099464}{\textit{Comm. Math. Phys.}}
 \textbf{174} (1995), 57--91,
 \href{http://arxiv.org/abs/hep-th/9405109}{arXiv:hep-th/9405109}.

\bibitem{Barnich:2000zw}
Barnich G., Brandt F., Henneaux M., Local {BRST} cohomology in gauge theories,
 \href{https://doi.org/10.1016/S0370-1573(00)00049-1}{\textit{Phys. Rep.}}
 \textbf{338} (2000), 439--569,
 \href{http://arxiv.org/abs/hep-th/0002245}{arXiv:hep-th/0002245}.

\bibitem{Barnich:2003wj}
Barnich G., Grigoriev M., Hamiltonian {BRST} and {B}atalin--{V}ilkovisky
 formalisms for second quantization of gauge theories,
 \href{https://doi.org/10.1007/s00220-004-1275-4}{\textit{Comm. Math. Phys.}}
 \textbf{254} (2005), 581--601,
 \href{http://arxiv.org/abs/hep-th/0310083}{arXiv:hep-th/0310083}.

\bibitem{Barnich:2006hbb}
Barnich G., Grigoriev M., {BRST} extension of the non-linear unfolded
 formalism, \textit{Bulg.~J.~Phys.} \textbf{33} (2006), no.~s1, 547--556,
 \href{http://arxiv.org/abs/hep-th/0504119}{arXiv:hep-th/0504119}.

\bibitem{Barnich:2009jy}
Barnich G., Grigoriev M., A~{P}oincar\'e lemma for sigma models of {AKSZ} type,
 \href{https://doi.org/10.1016/j.geomphys.2010.11.014}{\textit{J.~Geom.
 Phys.}} \textbf{61} (2011), 663--674,
 \href{http://arxiv.org/abs/0905.0547}{arXiv:0905.0547}.

\bibitem{Barnich:2010sw}
Barnich G., Grigoriev M., First order parent formulation for generic gauge
 field theories,
 \href{https://doi.org/10.1007/JHEP01(2011)122}{\textit{J.~High Energy Phys.}}
 \textbf{2011} (2011), no.~1, 122, 36~pages,
 \href{http://arxiv.org/abs/1009.0190}{arXiv:1009.0190}.

\bibitem{Barnich:2004cr}
Barnich G., Grigoriev M., Semikhatov A., Tipunin I., Parent field theory and
 unfolding in {BRST} first-quantized terms,
 \href{https://doi.org/10.1007/s00220-005-1408-4}{\textit{Comm. Math. Phys.}}
 \textbf{260} (2005), 147--181,
 \href{http://arxiv.org/abs/hep-th/0406192}{arXiv:hep-th/0406192}.

\bibitem{Batalin:2001fc}
Batalin I., Marnelius R., Superfield algorithms for topological field theories,
 in Multiple Facets of Quantization and Supersymmetry,
 \href{https://doi.org/10.1142/9789812777065_0021}{World Scientific
 Publishing}, River Edge, NJ, 2002, 233--251,
 \href{http://arxiv.org/abs/hep-th/0110140}{arXiv:hep-th/0110140}.

\bibitem{Batalin:1981jr}
Batalin I.A., Vilkovisky G.A., Gauge algebra and quantization,
 \href{https://doi.org/10.1016/0370-2693(81)90205-7}{\textit{Phys. Lett.~B}}
 \textbf{102} (1981), 27--31.

\bibitem{Batalin:1983wj}
Batalin I.A., Vilkovisky G.A., Feynman rules for reducible gauge theories,
 \href{https://doi.org/10.1016/0370-2693(83)90645-7}{\textit{Phys. Lett.~B}}
 \textbf{120} (1983), 166--170.

\bibitem{Bekaert:2013zya}
Bekaert X., Grigoriev M., Higher-order singletons, partially massless fields,
 and their boundary values in the ambient approach,
 \href{https://doi.org/10.1016/j.nuclphysb.2013.08.015}{\textit{Nuclear
 Phys.~B}} \textbf{876} (2013), 667--714,
 \href{http://arxiv.org/abs/1305.0162}{arXiv:1305.0162}.

\bibitem{Bekaert:2012vt}
Bekaert X., Grigoriev M., Notes on the ambient approach to boundary values of
 {A}d{S} gauge fields,
 \href{https://doi.org/10.1088/1751-8113/46/21/214008}{\textit{J.~Phys.~A}}
 \textbf{46} (2013), 214008, 23~pages,
 \href{http://arxiv.org/abs/1207.3439}{arXiv:1207.3439}.

\bibitem{Bekaert:2017bpy}
Bekaert X., Grigoriev M., Skvortsov E.D., Higher spin extension of
 {F}efferman--{G}raham construction,
 \href{http://doi.org/10.3390/universe4020017}{\textit{Universe}} \textbf{4}
 (2018), no.~2, 17, 26~pages,
 \href{http://arxiv.org/abs/1710.11463}{arXiv:1710.11463}.

\bibitem{Bocharov:1999}
Bocharov A.V., Chetverikov V.N., Duzhin S.V., Khor'kova N.G., Krasil'shchik
 I.S., Samokhin A.V., Torkhov~Yu.N., Verbovetsky A.M., Vinogradov A.M.,
 Symmetries and conservation laws for differential equations of mathematical
 physics, \textit{Transl. Math. Monogr.}, Vol.~182,
 \href{https://doi.org/10.1090/mmono/182}{American Mathematical Society},
 Providence, RI, 1999.

\bibitem{bonavolonta2013}
Bonavolont\`a G., Kotov A., On the space of super maps between smooth
 supermanifold, \href{http://arxiv.org/abs/1304.0394}{arXiv:1304.0394}.

\bibitem{Bonavolonta:2013mza}
Bonavolont\`a G., Kotov A., Local {BRST} cohomology for {AKSZ} field theories:
 {A}~global approach, in Mathematical Aspects of Quantum Field Theories, \textit{Math.
 Phys. Stud.}, \href{https://doi.org/10.1007/978-3-319-09949-1_10}{Springer},
 Cham, 2015, 325--341,
 \href{http://arxiv.org/abs/1310.0245}{arXiv:1310.0245}.

\bibitem{Bonavolonta:2012fh}
Bonavolont\`a G., Poncin N., On the category of {L}ie {$n$}-algebroids,
 \href{https://doi.org/10.1016/j.geomphys.2013.05.004}{\textit{J.~Geom.
 Phys.}} \textbf{73} (2013), 70--90,
 \href{http://arxiv.org/abs/1207.3590}{arXiv:1207.3590}.

\bibitem{Bonechi:2022aji}
Bonechi F., Cattaneo A.S., Zabzine M., Towards equivariant {Y}ang--{M}ills
 theory,
 \href{https://doi.org/10.1016/j.geomphys.2023.104836}{\textit{J.~Geom.
 Phys.}} \textbf{189} (2023), 104836, 17~pages,
 \href{http://arxiv.org/abs/2210.00372}{arXiv:2210.00372}.

\bibitem{Bonechi:2009kx}
Bonechi F., Mn\"ev P., Zabzine M., Finite-dimensional {AKSZ}-{BV} theories,
 \href{https://doi.org/10.1007/s11005-010-0423-3}{\textit{Lett. Math. Phys.}}
 \textbf{94} (2010), 197--228,
 \href{http://arxiv.org/abs/0903.0995}{arXiv:0903.0995}.

\bibitem{Boulanger:2015ova}
Boulanger N., Kessel P., Skvortsov E., Taronna M., Higher spin interactions in
 four-dimensions: {V}asiliev versus {F}ronsdal,
 \href{https://doi.org/10.1088/1751-8113/49/9/095402}{\textit{J.~Phys.~A}}
 \textbf{49} (2016), 095402, 52~pages,
 \href{http://arxiv.org/abs/1508.04139}{arXiv:1508.04139}.

\bibitem{Cattaneo:1999fm}
Cattaneo A.S., Felder G., A~path integral approach to the {K}ontsevich
 quantization formula,
 \href{https://doi.org/10.1007/s002200000229}{\textit{Comm. Math. Phys.}}
 \textbf{212} (2000), 591--611,
 \href{http://arxiv.org/abs/math.QA/9902090}{arXiv:math.QA/9902090}.

\bibitem{Cattaneo:2001ys}
Cattaneo A.S., Felder G., On the {AKSZ} formulation of the {P}oisson sigma
 model, \href{https://doi.org/10.1023/A:1010963926853}{\textit{Lett. Math.
 Phys.}} \textbf{56} (2001), 163--179.

\bibitem{Cattaneo:2012qu}
Cattaneo A.S., Mnev P., Reshetikhin N., Classical {BV} theories on manifolds
 with boundary, \href{https://doi.org/10.1007/s00220-014-2145-3}{\textit{Comm.
 Math. Phys.}} \textbf{332} (2014), 535--603,
 \href{http://arxiv.org/abs/1201.0290}{arXiv:1201.0290}.

\bibitem{Cattaneo:2015vsa}
Cattaneo A.S., Mnev P., Reshetikhin N., Perturbative quantum gauge theories on
 manifolds with boundary,
 \href{https://doi.org/10.1007/s00220-017-3031-6}{\textit{Comm. Math. Phys.}}
 \textbf{357} (2018), 631--730,
 \href{http://arxiv.org/abs/1507.01221}{arXiv:1507.01221}.

\bibitem{Chekmenev:2015kzf}
Chekmenev A., Grigoriev M., Boundary values of mixed-symmetry massless fields
 in {A}d{S} space,
 \href{http://doi.org/10.1016/j.nuclphysb.2016.10.006}{\textit{Nuclear Phys.
 B}} \textbf{913} (2016), 769--791,
 \href{http://arxiv.org/abs/1512.06443}{arXiv:1512.06443}.

\bibitem{costello2011renormalization}
Costello K., Renormalization and effective field theory, \textit{Math. Surveys
 Monogr.}, Vol.~170, \href{https://doi.org/10.1090/surv/170}{American
 Mathematical Society}, Providence, RI, 2011.

\bibitem{Didenko:2019xzz}
Didenko V.E., Gelfond O.A., Korybut A.V., Vasiliev M.A., Limiting shifted
 homotopy in higher-spin theory and spin-locality,
 \href{https://doi.org/10.1007/jhep12(2019)086}{\textit{J.~High Energy Phys.}}
 \textbf{2019} (2019), 086, 49~pages,
 \href{http://arxiv.org/abs/1909.04876}{arXiv:1909.04876}.

\bibitem{Dneprov:2022jyn}
Dneprov I., Grigoriev M., Presymplectic {BV-AKSZ} formulation of conformal
 gravity, \href{https://doi.org/10.1140/epjc/s10052-022-11082-6}{\textit{Eur.
 Phys.~J.~C}} \textbf{83} (2023), 6, 20~pages,
 \href{http://arxiv.org/abs/2208.02933}{arXiv:2208.02933}.

\bibitem{Dneprov:2024cvt}
Dneprov I., Grigoriev M., Gritzaenko V., Presymplectic minimal models of local
 gauge theories,
 \href{https://doi.org/10.1088/1751-8121/ad65a3}{\textit{J.~Phys.~A}}
 \textbf{57} (2024), 335402, 29~pages,
 \href{http://arxiv.org/abs/2402.03240}{arXiv:2402.03240}.

\bibitem{Gelfond:2019tac}
Gelfond O.A., Vasiliev M.A., Spin-locality of higher-spin theories and
 star-product functional classes,
 \href{https://doi.org/10.1007/jhep03(2020)002}{\textit{J.~High Energy Phys.}}
 \textbf{2020} (2020), no.~3, 002, 52~pages,
 \href{http://arxiv.org/abs/1910.00487}{arXiv:1910.00487}.

\bibitem{Gomis:1995he}
Gomis J., Par\'{\i}s J., Samuel S., Antibracket, antifields and gauge-theory
 quantization,
 \href{https://doi.org/10.1016/0370-1573(94)00112-G}{\textit{Phys. Rep.}}
 \textbf{259} (1995), 145,
 \href{http://arxiv.org/abs/hep-th/9412228}{arXiv:hep-th/9412228}.

\bibitem{Grigoriev:2006tt}
Grigoriev M., Off-shell gauge fields from {BRST} quantization,
 \href{http://arxiv.org/abs/hep-th/0605089}{arXiv:hep-th/0605089}.

\bibitem{Grigoriev:2016wmk}
Grigoriev M., Presymplectic structures and intrinsic {L}agrangians,
 \href{http://arxiv.org/abs/1606.07532}{arXiv:1606.07532}.

\bibitem{Grigoriev:2022zlq}
Grigoriev M., Presymplectic gauge {PDE}s and {L}agrangian {BV} formalism beyond
 jet-bundles, in The Diverse World of {PDE}s---Geometry and Mathematical
 Physics, \textit{Contemp. Math.}, Vol.~788,
 \href{https://doi.org/10.1090/conm/788/15822}{American Mathematical Society},
 Providence, RI, 2023, 111--133,
 \href{http://arxiv.org/abs/2212.11350}{arXiv:2212.11350}.

\bibitem{Grigoriev:1999qz}
Grigoriev M., Damgaard P.H., Superfield {BRST} charge and the master action,
 \href{https://doi.org/10.1016/S0370-2693(00)00050-2}{\textit{Phys. Lett.~B}}
 \textbf{474} (2000), 323--330,
 \href{http://arxiv.org/abs/hep-th/9911092}{arXiv:hep-th/9911092}.

\bibitem{Grigoriev:2019ojp}
Grigoriev M., Kotov A., Gauge {PDE} and {AKSZ}-type sigma models,
 \href{https://doi.org/10.1002/prop.201910007}{\textit{Fortschr. Phys.}}
 \textbf{67} (2019), 1910007, 13~pages,
 \href{http://arxiv.org/abs/1903.02820}{arXiv:1903.02820}.

\bibitem{Grigoriev:2020xec}
Grigoriev M., Kotov A., Presymplectic {AKSZ} formulation of {E}instein gravity,
 \href{https://doi.org/10.1007/jhep09(2021)181}{\textit{J.~High Energy Phys.}}
 \textbf{2021} (2021), no.~9, 181, 23~pages,
 \href{http://arxiv.org/abs/2008.11690}{arXiv:2008.11690}.

\bibitem{Grigoriev:2025vsl}
Grigoriev M., Mamekin A., Presymplectic {BV}-{AKSZ} for {$N=1$}, {$D=4$}
 supergravity, \href{http://arxiv.org/abs/2503.04559}{arXiv:2503.04559}.

\bibitem{Grigoriev:2023kkk}
Grigoriev M., Markov M., Asymptotic symmetries of gravity in the gauge {PDE}
 approach, \href{http://doi.org/10.1088/1361-6382/ad4ae0}{\textit{Classical
 Quantum Gravity}} \textbf{41} (2024), 135009, 27~pages,
 \href{http://arxiv.org/abs/2310.09637}{arXiv:2310.09637}.

\bibitem{Grigoriev:2023lcc}
Grigoriev M., Rudinsky D., Notes on the {$L_\infty$}-approach to local gauge
 field theories,
 \href{https://doi.org/10.1016/j.geomphys.2023.104863}{\textit{J.~Geom.
 Phys.}} \textbf{190} (2023), 104863, 21~pages,
 \href{http://arxiv.org/abs/2303.08990}{arXiv:2303.08990}.


\bibitem{Henneaux:1990rx}
Henneaux M., Spacetime locality of the {BRST} formalism,
 \href{http://doi.org/10.1007/BF02099287}{\textit{Comm. Math. Phys.}}
 \textbf{140} (1991), 1--13.

\bibitem{HT-book}
Henneaux M., Teitelboim C., Quantization of gauge systems,
 \href{https://doi.org/10.1515/9780691213866}{Princeton University Press},
 Princeton, NJ, 1992.

\bibitem{Ikeda:2012pv}
Ikeda N., Lectures on {AKSZ} sigma models for physicists, in Noncommutative
 Geometry and Physics.~4,
 \href{https://doi.org/10.1142/9789813144613_0003}{World Scientific
 Publishing}, Hackensack, NJ, 2017, 79--169,
 \href{http://arxiv.org/abs/1204.3714}{arXiv:1204.3714}.

\bibitem{Kazinski:2005eb}
Kazinski P.O., Lyakhovich S.L., Sharapov A.A., Lagrange structure and
 quantization,
 \href{https://doi.org/10.1088/1126-6708/2005/07/076}{\textit{J.~High Energy
 Phys.}} \textbf{2005} (2005), no.~7, 076, 42~pages,
 \href{http://arxiv.org/abs/hep-th/0506093}{arXiv:hep-th/0506093}.

\bibitem{Kobayashi:1962a}
Kobayashi E.T., A~remark on the {N}ijenhuis tensor,
 \href{https://doi.org/10.2140/pjm.1962.12.963}{\textit{Pacific~J.~Math.}}
 \textbf{12} (1962), 963--977, {E}rrarum,
 \href{https://doi.org/10.2140/pjm.1962.12.1467}{\textit{Pacific~J.~Math.}} \textbf{12}
 (1962), 1467.

\bibitem{Kontsevich:1997vb}
Kontsevich M., Deformation quantization of {P}oisson manifolds,
 \href{https://doi.org/10.1023/B:MATH.0000027508.00421.bf}{\textit{Lett. Math.
 Phys.}} \textbf{66} (2003), 157--216,
 \href{http://arxiv.org/abs/q-alg/9709040}{arXiv:q-alg/9709040}.

\bibitem{Kotov:2007nr}
Kotov A., Strobl T., Characteristic classes associated to {$Q$}-bundles,
 \href{https://doi.org/10.1142/S0219887815500061}{\textit{Int.~J.~Geom.
 Methods Mod. Phys.}} \textbf{12} (2015), 1550006, 26~pages,
 \href{http://arxiv.org/abs/0711.4106}{arXiv:0711.4106}.

\bibitem{Krasil?shchik-Lychagin-Vinogradov}
Krasil'shchik I.S., Lychagin V.V., Vinogradov A.M., Geometry of jet spaces and
 nonlinear partial differential equations, \textit{Adv. Stud. Contemp. Math},
 Vol.~1, Gordon and Breach Science Publishers, New York, 1986.

\bibitem{Krasil'shchik:2010ij}
Krasil'shchik J., Verbovetsky A., Geometry of jet spaces and integrable
 systems,
 \href{https://doi.org/10.1016/j.geomphys.2010.10.012}{\textit{J.~Geom.
 Phys.}} \textbf{61} (2011), 1633--1674,
 \href{http://arxiv.org/abs/1002.0077}{arXiv:1002.0077}.

\bibitem{Lee2013}
Lee J.M., Introduction to smooth manifolds, 2nd ed., \textit{Grad. Texts Math.}, Vol.~218, \href{https://doi.org/10.1007/978-1-4419-9982-5}{Springer}, New
 York, 2012.

\bibitem{Lyakhovich:2007cw}
Lyakhovich S.L., Sharapov A.A., Quantization of {D}onaldson--{U}hlenbeck--{Y}au
 theory, \href{https://doi.org/10.1016/j.physletb.2007.09.029}{\textit{Phys.
 Lett.~B}} \textbf{656} (2007), 265--271,
 \href{http://arxiv.org/abs/0705.1871}{arXiv:0705.1871}.

\bibitem{Mehta:2007rgt}
Mehta R.A., {$Q$}-algebroids and their cohomology,
 \href{https://doi.org/10.4310/jsg.2009.v7.n3.a1}{\textit{J.~Symplectic
 Geom.}} \textbf{7} (2009), 263--293,
 \href{http://arxiv.org/abs/math.DG/0703234}{arXiv:math.DG/0703234}.

\bibitem{Misuna:2024dlx}
Misuna N., Unfolded formulation of {$4 d$} {Y}ang--{M}ills theory,
 \href{https://doi.org/10.1016/j.physletb.2025.139882}{\textit{Phys. Lett.~B}}
 \textbf{870} (2025), 139882, 6~pages,
 \href{http://arxiv.org/abs/2408.13212}{arXiv:2408.13212}.

\bibitem{Piguet:1995er}
Piguet O., Sorella S.P., Algebraic renormalization. Perturbative
 renormalization, symmetries and anomalies, \textit{Lecture Notes in Phys. New
 Ser.~m Monogr.}, Vol.~28,
 \href{https://doi.org/10.1007/978-3-540-49192-7}{Springer}, Berlin, 1995.

\bibitem{Roytenberg:2002nu}
Roytenberg D., On the structure of graded symplectic supermanifolds and
 {C}ourant algebroids, in Quantization, {P}oisson Brackets and Beyond
 ({M}anchester, 2001), \textit{Contemp. Math.}, Vol.~315,
 \href{https://doi.org/10.1090/conm/315/05479}{American Mathematical Society},
 Providence, RI, 2002, 169--185,
 \href{http://arxiv.org/abs/math.SG/0203110}{arXiv:math.SG/0203110}.

\bibitem{Roytenberg:2006qz}
Roytenberg D., {AKSZ}-{BV} formalism and {C}ourant algebroid-induced
 topological field theories,
 \href{https://doi.org/10.1007/s11005-006-0134-y}{\textit{Lett. Math. Phys.}}
 \textbf{79} (2007), 143--159,
 \href{http://arxiv.org/abs/hep-th/0608150}{arXiv:hep-th/0608150}.

\bibitem{Schwarz:1992nx}
Schwarz A., Geometry of {B}atalin--{V}ilkovisky quantization,
 \href{https://doi.org/10.1007/BF02097392}{\textit{Comm. Math. Phys.}}
 \textbf{155} (1993), 249--260,
 \href{http://arxiv.org/abs/hep-th/9205088}{arXiv:hep-th/9205088}.

\bibitem{Severa:2001tze}
Severa P., Some title containing the words ``homotopy'' and ``symplectic'',
 e.g. this one, in Travaux Math\'ematiques. {F}asc.~{XVI}, \textit{Trav.
 Math.}, Vol.~16, Universit\'e du Luxembourg, Luxembourg, 2005, 121--137,
 \href{http://arxiv.org/abs/math.SG/0105080}{arXiv:math.SG/0105080}.

\bibitem{Sharapov:2021drr}
Sharapov A., Skvortsov E., Higher spin gravities and presymplectic {AKSZ}
 models,
 \href{https://doi.org/10.1016/j.nuclphysb.2021.115551}{\textit{Nuclear
 Phys.~B}} \textbf{972} (2021), 115551, 56~pages,
 \href{http://arxiv.org/abs/2102.02253}{arXiv:2102.02253}.

\bibitem{Sharapov:2022faa}
Sharapov A., Skvortsov E., Sukhanov A., Van~Dongen R., Minimal model of chiral
 higher spin gravity,
 \href{https://doi.org/10.1007/jhep09(2022)134}{\textit{J.~High Energy Phys.}}
 \textbf{2022} (2022), no.~9, 134, 31~pages,
 \href{http://arxiv.org/abs/2205.07794}{arXiv:2205.07794}.

\bibitem{Sharapov:2022wpz}
Sharapov A., Skvortsov E., Van~Dongen R., Chiral higher spin gravity and convex
 geometry,
 \href{https://doi.org/10.21468/scipostphys.14.6.162}{\textit{SciPost Phys.}}
 \textbf{14} (2023), 162, 15~pages,
 \href{http://arxiv.org/abs/2209.01796}{arXiv:2209.01796}.

\bibitem{Sharapov:2016sgx}
Sharapov A.A., Variational tricomplex, global symmetries and conservation laws
 of gauge systems,
 \href{https://doi.org/10.3842/SIGMA.2016.098}{\textit{SIGMA}} \textbf{12}
 (2016), 098, 24~pages,
 \href{http://arxiv.org/abs/1607.01626}{arXiv:1607.01626}.

\bibitem{Skvortsov:2015lja}
Skvortsov E., Taronna M., On locality, holography and unfolding,
 \href{https://doi.org/10.1007/JHEP11(2015)044}{\textit{J.~High Energy Phys.}}
 \textbf{2015} (2015), no.~11, 044, 36~pages,
 \href{http://arxiv.org/abs/1508.04764}{arXiv:1508.04764}.

\bibitem{Vaintrob:1997}
Vaintrob A.Yu., Lie algebroids and homological vector fields,
 \href{https://doi.org/10.1070/RM1997v052n02ABEH001802}{\textit{Russian Math.
 Surveys}} \textbf{52} (1997), 428--429.

\bibitem{Vasiliev:1988xc}
Vasiliev M.A., Equations of motion of interacting massless fields of all spins
 as a~free differential algebra,
 \href{https://doi.org/10.1016/0370-2693(88)91179-3}{\textit{Phys. Lett.~B}}
 \textbf{209} (1988), 491--497.

\bibitem{Vasiliev:1999ba}
Vasiliev M.A., Higher spin gauge theories: {S}tar-product and {AdS} space, in
 The Many Faces of the Superworld,
 \href{https://doi.org/10.1142/9789812793850_0030}{World Scientific
 Publishing}, 2000, 533--610,
 \href{http://arxiv.org/abs/hep-th/9910096}{arXiv:hep-th/9910096}.

\bibitem{Vasiliev:2003ev}
Vasiliev M.A., Nonlinear equations for symmetric massless higher spin fields in
 {$({\rm A}){\rm dS}_d$},
 \href{https://doi.org/10.1016/S0370-2693(03)00872-4}{\textit{Phys. Lett.~B}}
 \textbf{567} (2003), 139--151,
 \href{http://arxiv.org/abs/hep-th/0304049}{arXiv:hep-th/0304049}.

\bibitem{Vasiliev:2005zu}
Vasiliev M.A., Actions, charges and off-shell fields in the unfolded dynamics
 approach,
 \href{https://doi.org/10.1142/S0219887806001016}{\textit{Int.~J.~Geom.
 Methods Mod. Phys.}} \textbf{3} (2006), 37--80,
 \href{http://arxiv.org/abs/hep-th/0504090}{arXiv:hep-th/0504090}.

\bibitem{Vinogradov1981}
Vinogradov A.M., Geometry of nonlinear differential equations,
 \href{http://doi.org/10.1007/BF01084594}{\textit{J.~Math. Sci.}} \textbf{17}
 (1981), 1624--1649.

\bibitem{Voronov:2009nr}
Voronov T.T., On a~non-abelian {P}oincar\'e lemma,
 \href{https://doi.org/10.1090/S0002-9939-2011-11116-X}{\textit{Proc. Amer.
 Math. Soc.}} \textbf{140} (2012), 2855--2872,
 \href{http://arxiv.org/abs/0905.0287}{arXiv:0905.0287}.

\end{thebibliography}

\pdfbookmark[1]{References}{ref}
\LastPageEnding

\end{document}